# Quantum measurements and new concepts for experiments with trapped ions

Ch. Wunderlich and Ch. Balzer

*Institut für Laser-Physik, Universität Hamburg,*
*Luruper Chaussee 149, 22761 Hamburg, Germany*

25 November 2002

# Contents

# 1  Overview

Quantum mechanics is a tremendously successful theory playing a central role in natural sciences even beyond physics, and has been verified in countless experiments, some of which were carried out with very high precision. Despite its great success and its history reaching back more than hundred years, still today the interpretation of quantum mechanics challenges our intuition that has been formed by an environment governed by classical physical laws.

Quantum optical experiments may come very close to idealized situations of gedanken experiments originally conceived to test and better understand the predictions and implications of quantum theory. An experimental system ideally suited to carry out such experiments will be dealt with in this work: electrodynamically trapped ions provide us with individual localized quantum systems well isolated from the environment. The interaction with electromagnetic radiation allows for preparation and detection of quantum states, even of single ions (Neuhauser et al. 1980). Since the first storage and detection of a collection of ions in Paul and Penning traps has been reported (Fischer 1959, Church & Dehmelt 1969, Ifflander & Werth 1977), a large variety of intriguing experiments were carried out, for instance, the demonstration of optical cooling (Neuhauser et al. 1978, Wineland et al. 1978) and experiments related to fundamental physical questions (for instance, (Sauter et al. 1986, Bergquist et al. 1986, Diedrich & Walther 1987, Schubert et al. 1992, Howe et al. 2001, Guthöhrlein et al. 2001).) Also, for precision measurements and frequency standards the use of trapped ions is well established (for instance, (Stenger et al. 2001, Diddams et al. 2001, Becker et al. 2001).)

The fact that quantum mechanics makes only statistical predictions let Albert Einstein and others doubt whether this theory is correct, or more specific, whether it gives a complete description of physical reality as they perceived it. Einstein cast part of his doubts about this theory in the words "Gott würfelt nicht" ("God doesn't roll dice",) that is, according to his opinion laws of nature do *not* contain this intrinsic randomness and a proper theory should account for that.

Another puzzling feature of quantum mechanics was pointed out by Einstein, Podolsky, and Rosen (EPR) in (Einstein et al. 1935). Quantum theory predicts correlations between two or more quantum systems once an entangled state of these systems has been generated. These correlations persist even after the quantum systems have been brought to spacelike separated points. The statistical nature of quantum mechanical predictions, and the superposition principle, together with quantum mechanical commutation relations give rise to such nonlocal correlations (Einstein et al. 1935). Einstein found this, what he later called "spukhafte Fernwirkung" ("spooky action at a distance") deeply disturbing and concluded that quantum mechanics is an incomplete theory. The term "Verschränkung" (entanglement) has been coined by E. Schrödinger to describe such correlated quantum systems (Schrödinger 1935). Recently, entangled states of various physical systems have been created and analyzed in experiments (a review can be found in (Whitaker 2000).) All experimental findings have been in agreement with quantum mechanical predictions.

There is no *a priori* reason not to apply quantum mechanics to objects like a measurement apparatus made up from a large number of elementary constituents each of which is perfectly described by quantum theory. This, however, may be a cause for yet more discomfort, since it leads to seemingly paradoxical or absurd consequences as Erwin Schrödinger pointed out (Schrödinger 1935). With a gedanken experiment he illustrated the consequences of including an object usually described by classical physics (he chose a cat) into a quantum mechanical description [1]. The cat is 'coupled' to a quantum system prepared in a superposition state, and in the course of the gedanken experiment the cat, too, assumes a superposition state of 'being dead' and 'being alive' (Schrödinger 1935): an entangled state of quantum system and cat results.

The cat can be viewed as a macroscopic apparatus that is used to measure the state of a quantum system. Thus, if the quantum system initially is in a superposition of two states, then linearity of quantum mechanics demands the measurement apparatus, too, to be in a superposition of two of its meter states. This is clearly not what we usually observe in experiments. Reference (Brune et al. 1996) describes a cavity QED experiment where an electromagnetic field acts as meter for the quantum state of individual atoms. It is shown how the decay of the initially prepared superposition of meter states is the faster the larger the initial separation of these states is. For macroscopically distinct meter states this decay of a superposition state into a statistical mixture of states (that is, either one *or* the other is realized) is usually too fast to be observable experimentally. Thus, superpositions of macroscopically distinct states are never observed. Schrödinger-cat like states have also been investigated with trapped ions (Myatt et al. 2000) and superconducting quantum interference devices (Friedman et al. 2000).

The first step in a measurement process requires some interaction between the quantum system and a second system (the probe), and consequently a correlation is established between the two systems (In general, this will result in an entangled state between quantum system and probe.) This correlation reduces or even destroys the quantum system's capability to display characteristics of a superposition state in subsequent local operations, and the appropriate description of the quantum system alone is a statistical mixture of states. The coupling of the probe to a macroscopic apparatus leads to a reduction of the probe itself from a coherent superposition into a statistical mixture (for instance, (Zurek 1991, Giulini et al. 1996) and references therein.) When the apparatus is finally found in one of its meter states, quantum mechanics tells us that the quantum system is reset to the state correlated with this particular meter state. This will be evident in any subsequent manipulation the quantum system is subjected to.

If the quantum system would undergo some kind of evolution as long as it is not being measured, then the measurement process might impede or even freeze this evolution. This slowing down (or coming to a complete halt) of the dynamics of a quantum system when subjected to frequent measurements (von Neumann 1932) has been termed quantum Zeno effect or quantum Zeno paradox (Misra & Sudarshan 1977).

An unambiguous demonstration of this effect requires measurements on individual quantum systems as opposed to ensemble measurements. Such an experiment has been carried out with individual electrodynamically trapped $Yb^+$ ions prepared in a well defined quantum state, and it is shown that even

---
[1] Arguably classical physics is not sufficient to describe a cat. For the purpose of the gedanken experiment, therefore, it might be useful to choose an inanimate macroscopic object

negative-result measurements (which do not involve local interaction between quantum system and apparatus in a "classical" sense) impede the quantum system's evolution (section 4.)

Now we turn to the concept of a quantum state that is a central ingredient of quantum theory. How can an arbitrary *unknown* state of a quantum system be determined accurately? The determination of the set of expectation values of the observables associated with a specific quantum state is complicated by the fact that after a measurement of one observable, information on the complementary observable is no longer available. Only if infinitely many identical copies of a given state were available could this task be achieved. Since this requirement cannot be fulfilled in experiments, it is of interest to investigate ways to gain optimal knowledge of a given quantum state making use of *finite* resources. In addition, quantum state estimation is, for instance, relevant for quantum communication where quantum information at the receiver end of a quantum channel has to be deciphered.

If $N$ identically prepared quantum systems in an unknown arbitrary state are available, how can this state be determined? In other words, what is the optimal strategy to gain the maximal amount of information about the state of a quantum system using finite physical resources? Quantum states of various physical systems such as light fields, molecular wave packets, motional states of trapped ions and atomic beams have been determined experimentally (for a review of recent work see, for instance, (Schleich &Raymer 1997, Freyberger et al. 1997, Bužek et al. 1998, Walmsley & Waxer 1998, White et al. 1999, Lvovsky et al. 2001).)

Optimal strategies to read out information encoded in the quantum state of a given number $N$ of identical *two-state* systems (qubits) have been proposed in recent years. However, they require intricate measurements using a basis of entangled states. It is desirable to have a measurement strategy at hand that gives an estimate of a quantum state with high fidelity, even if $N$ measurements are performed separately (even sequentially) on each individual qubit, that is, if a factorizing basis is employed for state estimation. Sequential measurements on arbitrary but identically prepared states of a qubit, the ground state hyperfine levels of electrodynamically trapped $^{171}$Yb$^+$, are described in section 5. The measurement basis is varied during a sequence of $N$ measurements conditioned on the results of previous measurements in this sequence. The experimental efficiency and fidelity of such a self-learning measurement (Fischer et al. 2000) is compared with strategies where the measurement basis is randomly chosen during a sequence of $N$ measurements.

In addition to puzzling us with fundamental questions regarding, for example, the measurement process, quantum mechanics holds the opportunity to put its laws to practical use. In the field of quantum information processing (QIP) and communication basic elements of computers are explored that would be able to solve problems that, for all practical purposes, cannot be handled by classical computers and communication devices ((Feynman 1982, Deutsch 1985, Gruska 1999, Nielsen & Chuang 2000), and references therein.) The computation of properties of quantum systems themselves is particularly suited to be performed on a quantum computer, even on a device where logic operations can only be carried out with limited precision. Exchange of information can be made secure by using encrypting methods that rely on quantum properties, for instance, of optical radiation. While exploring these routes to new types of computing and communication, again much will be learned about still unsolved issues in quantum mechanics, for instance, regarding the characterization of entanglement (Lewenstein et al. 2000). The experimental system described in this work is well suited to conduct investigations in this new field.

The great potential that trapped ions have as a physical system for quantum information processing (QIP) was first recognized in (Cirac & Zoller 1995), and important experimental steps have been undertaken towards the realization of an elementary quantum computer with this system (for instance, (Wineland et al. 1998, Appasamy et al. 1998, Roos et al. 1999, Hannemann et al. 2002).) At the same time, the advanced state of experiments with trapped ions reveals the difficulties that still have to be overcome.

Using $^{171}$Yb$^+$ ions we have realized a quantum channel, that is, propagation of quantum information in time or space, under the influence of well controlled disturbances. The parameters characterizing the quantum channel can be adjusted at will and various types of quantum channels (that may occur in other experimental systems, too) can be implemented with individual ions. Thus a model system is realized to investigate, for example, the reconstruction of quantum information after transmission through a noisy quantum channel (section 6.1.) Transfer of quantum states becomes important when quantum information is distributed between different quantum processors, as is envisaged, for instance, for ion trap quantum information processing (Pellizzari 1997, van Enk et al. 1999). Furthermore, codes for quantum information processing, and in particular error correction codes may be tested for their applicability under well defined, non-ideal conditions.

These experiments demonstrate the ability to prepare arbitrary states of this SU(2) system with very high precision – a prerequisite for quantum information processing. The coherence time of the hyperfine qubit in $^{171}$Yb$^+$ is long on the time scale of qubit operations and is essentially limited by the coherence time of microwave radiation used to drive the qubit transition.

In addition to the ability to perform arbitrary single-qubit operations, a second fundamental type of operation is required for QIP: conditional quantum dynamics with, at least, two qubits. Any quantum algorithm can then be synthesized using these elementary building blocks (DiVincenzo 1995, Barenco et al. 1995). While two internal states of each trapped ion serve as a qubit, communication between these qubits, necessary for conditional dynamics, is achieved via the vibrational motion of the ion string in a linear trap (the "bus -qubit") (Cirac & Zoller 1995). Thus, it is necessary to couple external (motional) and internal degrees of freedom. Common to all experiments performed to date – related either to QIP or other research fields – that require some kind of coupling between internal and external degrees of freedom of atoms is the use of *optical* radiation for this purpose. The recoil energy $E_r = (\hbar k)^2/2m$ taken up by an atom upon absorption or emission of a photon may change the atom's motional state ($k = 2\pi/\lambda$, $\lambda$ is the wavelength of the applied electromagnetic radiation, and $m$ is the mass of the ion.) In order for this to happen with appreciable probability with a harmonically trapped atom, the ratio between $E_r$ and the quantized motional energy of the trapped atom, $\hbar\nu$ should not be too small ($\nu$ is the angular frequency of the vibrational mode to be excited.) Therefore, in usual traps, driving radiation in the optical regime is necessary to couple internal and external dynamics of trapped atoms.

The distance between neighboring ions $\delta z$ in a linear electrodynamic ion trap is determined by the mutual Coulomb repulsion of the ions and the time averaged force exerted on the ions by the electrodynamic trapping field. Manipulation of individual ions is usually achieved by focusing electromagnetic radiation to a spot size much smaller than $\delta z$. Again, only optical radiation is useful for this purpose.

In section 6.2 a new concept for ion traps is described that allows for experiments requiring individual addressing of ions and conditional dynamics with several ions even with radiation in the *radio frequency* (rf) or *microwave* (mw) regime. It is shown how an additional magnetic field gradient applied to an electrodynamic trap individually shifts ionic qubit resonances making them distinguishable in frequency space. Thus, individual addressing for the purpose of single qubit operations becomes possible using long-wavelength radiation. At the same time, a coupling term between internal and motional states arises even when rf or mw radiation is applied to drive qubit transitions. Thus, conditional quantum dynamics can be carried out in this modified electrodynamic trap, and in such a new type of trap all schemes originally devised for *optical* QIP in ion traps can be applied in the rf or mw regime, too.

Many phenomena that were only recently studied in the optical domain form the basis for techniques belonging to the standard repertoire of coherent manipulation of nuclear and electronic magnetic moments associated with their spins. Nuclear magnetic resonance (NMR) experiments have been tremendously successful in the field of QIP taking advantage of highly sophisticated experimental techniques. However, NMR experiments usually work with macroscopic ensembles of spins and considerable effort has to be devoted to the preparation of pseudo-pure states of spins with initial thermal population distribution. This preparation leads to an exponentially growing cost (with the number $N$ of qubits) either in signal strength or the number of experiments involved (Vandersypen et al. 2000), since the fraction of spins in their ground state is proportional to $N/2^N$.

Trapped ions, on the other hand, provide individual qubits – for example, hyperfine states as described in this work – well isolated from their environment with read-out efficiency near unity. It would be desirable to combine the advantages of trapped ions and NMR techniques in future experiments using either "conventional" ion trap methods, but now with mw radiation as outlined above, or, as described in the second part of section 6.2.2, treating the ion string as a $N$-qubit molecule with adjustable spin-spin coupling constants: In a suitably modified ion trap, ionic qubit states are pairwise coupled. This spin-spin coupling can be formally described in the same way as J-coupling in molecules used for NMR, even though the physical origin of the interaction is very different. Thus, successful techniques and technology developed in spin resonance experiments, like NMR or ESR, can immediately be applied to trapped ions. An advantage of an artificial "molecule" in a trap is that the coupling constants $J_{ij}$ between qubits $i$ and $j$ can be chosen by the experimenter by setting the magnetic field gradient, the secular trap frequency, and the type of ions used. In addition, *individual* spins can be detected state selectively with an efficiency close to 100% by collecting scattered resonance fluorescence.

Another avenue towards quantum computation with trapped ions is the use of an electric quadrupole transition (E2 transition) as a qubit (Appasamy et al. 1998, Schmidt-Kaler et al. 2000, Barton et al. 2000, Hughes et al. 1998). Section 6.3 gives an account of experiments carried out with $Ba^+$ and $^{172}Yb^+$ ions where the E2 transition between the ground state $S_{1/2}$ and the metastable excited $D_{5/2}$ state is investigated.

Cooling of the collective motion of several particles is prerequisite for implementing conditional quantum dynamics on trapped ions. A study of the collective vibrational motion of two trapped $^{138}Ba^+$ ions cooled by two light fields is described in section 6.3.3. Parameter regimes of the laser light irradiating the ions can be identified that imply most efficient laser cooling and are least susceptible to drifts, fluctuations, and uncertainties in laser parameters (Reißet al. 2002).

# 2 Spin resonance with single $Yb^+$ ions

In this section we introduce experiments with $^{171}Yb^+$ ions demonstrating the precise manipulation of hyperfine states of single ions essentially free of longitudinal and transverse relaxation. The experimental techniques outlined here, form the basis for further experiments with individual $Yb^+$ ions described in sections 4, 5, and 6.

## 2.1 Experimental setup for $Yb^+$

$^{171}Yb^+$ or $^{172}Yb^+$ ions are confined in a miniature Paul trap (diameter of 2 mm). Excitation of the $S_{1/2}$ - $P_{1/2}$ transition of $Yb^+$ serves for initial cooling and detection of resonantly scattered light near 369nm (Figure 1). For this purpose, infrared light near 738nm is generated by a laser system based on a commercial Titanium:Sapphire laser and frequency doubled using a $LiIO_3$ crystal mounted at the center of a homemade ring resonator. The emission frequency is stabilized against drift using an additional reference resonator.

Optical pumping into the $D_{3/2}$ state is prevented by illuminating the ions with laser light near 935nm. This couples state $|D_{3/2}, F=1\rangle$ via a dipole allowed transition to state $|[3/2]_{1/2},F=0\rangle$ that in turn decays to the ground state $|S_{1/2}, F=1\rangle$. Light near 935nm is produced by a homemade tunable, stabilized diode laser. Excitation spectra recorded with this laser have been recorded that exhibit sidebands due to micromotion of an ion in the trap. Making these sidebands disappear by adjusting the voltages applied to additional electrodes close to the trap serves for positioning the ion the field free potential minimum at the center of the trap.

The quantum mechanical two-state system used for the experiments described in sections 4, 5, and 6 is the $S_{1/2}$ ground-state hyperfine doublet with total angular momentum $F = 0, 1$ of $^{171}Yb^+$. The

$$|0\rangle \equiv |S_{1/2}, F = 0\rangle \leftrightarrow |S_{1/2}, F = 1, m_F = 0\rangle \equiv |1\rangle \qquad (1)$$

transition with Bohr frequency $\omega_0$ is driven by a quasiresonant microwave (mw) field with angular frequency near $\omega = 2\pi\,12.6$ GHz. The time evolution of the system is virtually free of decoherence, that is, transversal and longitudinal relaxation rates are negligible. However, imperfect preparation and detection limits the purity of the states. Photon-counting resonance fluorescence on the $S_{1/2}(F=1) \leftrightarrow P_{1/2}(F=0)$ transition at 369 nm serves for state selective detection with efficiency $> 98\%$. Optical pumping into the $|F = 1, m_F = \pm 1\rangle$ levels during a detection period is avoided when the $E$ vector of the linearly polarized light subtends $45^o$ with the direction of the applied dc magnetic field. The light is usually detuned to the red side of the resonance line by a few MHz in order to laser-cool the ion. Cooling is achieved by simultaneously irradiating the ion with light from both laser sources and with microwave radiation.

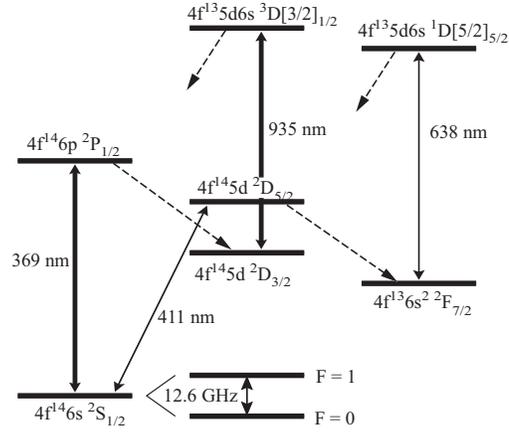

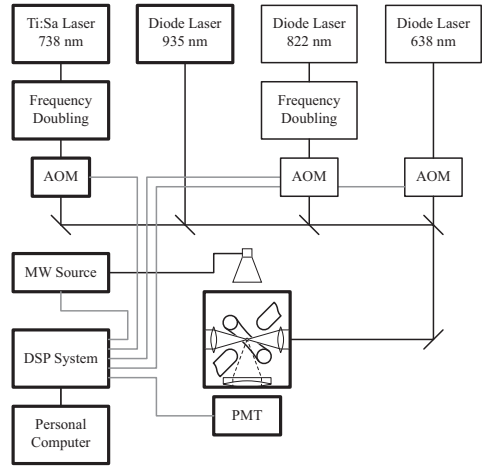

Figure 1: Top: Relevant energy levels of $^{171}$Yb$^+$ . The hyperfine splitting is shown only for the ground state (not to scale.) Bottom: Schematic drawing of major experimental elements. All lasers are frequency stabilized employing reference resonators (not shown.) MW: microwave; PMT: photo multiplier tube; DSP: digital signal processing; AOM: acousto optic modulator. For most experiments described in this work (using $^{171}$Yb$^+$ ) the elements drawn with bold lines are used.

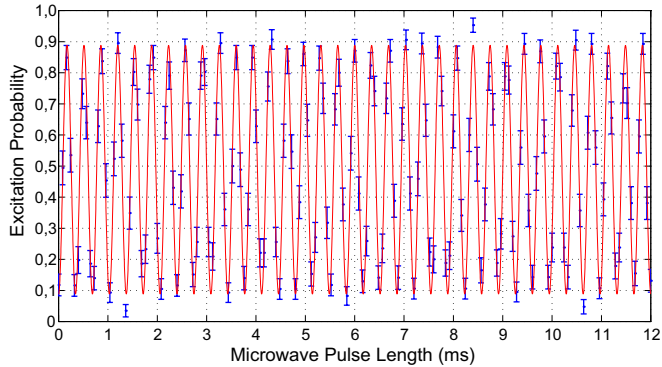

Figure 2: Rabi oscillations: Excitation probability of state $|1\rangle \equiv |S_{1/2} F = 0, m_F = 0\rangle$ of a single Yb$^+$ ion averaged over 85 preparation-detection cycles as a function of mw pulse length $t_{\mathrm{mw}}$. The solid line results from a fit using equation 4 giving $\Omega_R = 2.9165 \times 2\pi$kHz. The error bars indicate one standard deviation of the statistical error resulting from the finite number of preparation-detection cycles. The sub-unity contrast of the signal is due to imperfect initial state preparation by optical pumping (which will be improved in future experiments.)

When exciting the electric quadrupole transition $S_{1/2}$ - $D_{5/2}$ (section 6.3,) the Yb$^+$ ion may decay into the extremely long-lived $F_{7/2}$ state. Light generated by a tunable diode laser near 638nm resonantly couples this state to the excited state $D[5/2]_{5/2}$ such that optical pumping is avoided. The time needed to repump the ion from the $F_{7/2}$ state to the $S_{1/2}$ state has been determined as a function of the intensity of the laser light near 638nm (Riebe 2000). It saturates at $\approx$ 9ms.

## 2.2 Ground state hyperfine transition in $^{171}$Yb$^+$

The two hyperfine states of Yb$^+$ , $|0\rangle$ and $|1\rangle$ are coupled by a resonant, linearly polarized microwave field coherently driving transitions on this resonance. In a semiclassical description of the magnetic dipole interaction between a microwave field travelling in the $y - direction$ and the hyperfine states of $^{171}$Yb$^+$ the Hamiltonian reads

$$\begin{aligned} H &= \frac{\hbar}{2}\omega_0\sigma_z - \vec{\mu} \cdot \vec{B} \\ &= \frac{\hbar}{2}\omega_0\sigma_z + \frac{\hbar}{2}\gamma B_x \cos(ky - \omega t + \phi')\sigma_x \end{aligned} \qquad (2)$$

where $\vec{\mu}$ is the magnetic dipole operator of the ion, $\vec{B} = (B_x \cos(ky - \omega t + \phi'), 0, 0)^T$ is the magnetic field associated with the microwave radiation, and $\gamma$ is the gyromagnetic ratio. The initial phase of the mw field, $\phi = ky + \phi'$ at the location of the ion is set to zero in what follows. Transforming this Hamiltonian according to $\tilde{H} = \exp(i(\omega/2)t\sigma_z) H \exp(-i(\omega/2)t\sigma_z)$, and invoking the rotating wave approximation yields the time evolution operator

$$U(t) = \exp\left[-\frac{i}{2}t\left(\delta\sigma_z + \Omega\sigma_x\right)\right] \quad (3)$$

governing the dynamics of the two-state system. The detuning $\delta \equiv \omega_0 - \omega$, the Rabi frequency is denoted by $\Omega = \gamma B_x/2$, and $\sigma_{z,x}$ represent the usual Pauli matrices. If the ion is initially prepared in state $|0\rangle$, then the probability to find it in state $|1\rangle$ after time $t$ is

$$P_1(t) = \left(\frac{\Omega}{\Omega_R}\right)^2 \sin^2\left(\frac{\Omega_R}{2}t\right) \quad (4)$$

where $\Omega_R \equiv \sqrt{\Omega^2 + \delta^2}$. A pure state $|\theta, \phi\rangle = \cos\frac{\theta}{2}|0\rangle + \sin\frac{\theta}{2}e^{i\phi}|1\rangle$ represented by a unit vector in 3D configuration space (Bloch vector) is prepared by driving the hyperfine doublet with mw pulses with appropriately chosen detuning $\delta \equiv \omega_0 - \omega$, and duration $t_{\text{mw}} = \theta/\Omega$, and by allowing for free precession for a prescribed time $t_{\text{p}} = \phi/\delta$.

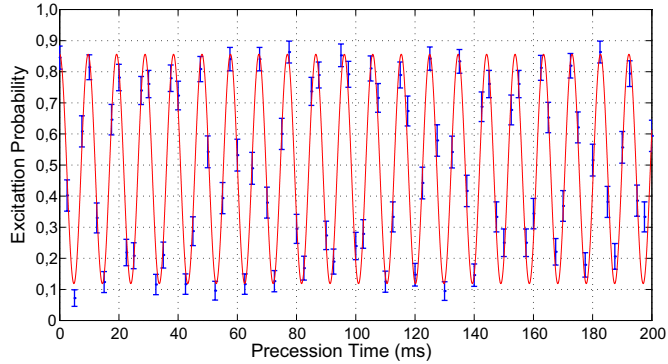

Figure 3: Data from a Ramsey-type experiment where the ion undergoes free precession between two subsequent mw $\pi/2$-pulses (detuning $\delta = 103.9 \times 2\pi$Hz, averaged over 100 realizations.) The error bars are of statistical origin. This experimental signal, too, is essentially free of decoherence and the contrast of the so-called Ramsey fringes is only limited by the finite preparation efficiency. The data displayed in Figure 2 and in this Figure show that single-qubit operations are carried out with high precision, an important prerequisite for scalable quantum computing.

The vertical bars in Fig. 2 indicate the experimentally determined excitation probability of state $|1\rangle$ (single $^{171}$Yb$^+$ ion) as a function of the mw pulse length $t_{\text{mw}}$; the solid line is a fit using equation 4 (Rabi oscillations.) The observed Rabi oscillations are free of decoherence over experimentally relevant time scales. However, the contrast of the oscillations is below unity, since the initial state $|0\rangle$ was prepared with probability 0.89. This limitation will be addressed in future experiments. Figure 3 displays data from a Ramsey-type experiment (Ramsey 1956) where the ion undergoes free precession for time $t_\text{p}$ between two subsequent mw pulses. This experimental signal, too, is essentially free of decoherence, and the contrast of the Ramsey fringes is only limited by the finite preparation efficiency. The data in Figure 2 and 3 show that single-qubit operations are carried out with high precision, an important prerequisite for scalable quantum computing.

# 3 Elements of quantum measurements

## 3.1 Measurements and Decoherence

In what follows, we consider the process of performing a measurement on a quantum system. We start by considering the interaction between the quantum system to be measured and a second system, the quantum probe, assuming that pure states of of both are prepared before an interaction between the two takes place. Initially, the state of the (unknown) quantum system $|\psi_i\rangle = \sum_n c_n |n\rangle$ ($|n\rangle$ are the eigenstates of the system Hamiltonian with complex coefficients $c_n$) and of the (known) state of the quantum probe $|\phi_i\rangle$ factorizes, that is we have $|\psi_i\rangle \otimes |\phi_i\rangle$. The interaction between system and probe is assumed to be governed by a Hamiltonian of the form ((Giulini et al. 1996) chapter 3)

$$H_\text{int} = \sum_n |n\rangle \langle n| \otimes \hat{A}_n \tag{5}$$

where $\hat{A}_n$ are operators acting only in the Hilbert space of the probe. They transform the probe conditioned on the state of the quantum system. If $|\psi_i\rangle = |n\rangle$, then, after the interaction has taken place the combined state of system and probe reads

$$|n\rangle |\phi_i\rangle \xrightarrow{H_\text{int}} |n\rangle |\phi_n\rangle \ . \tag{6}$$

For the sake of a clearer discussion in the following paragraphs we assume that $\langle \phi_k | \phi_l \rangle = \delta_{kl}$. In general, if the quantum system is initially prepared in a superposition state, the first step of the measurement will result in an entangled state between system and probe

$$|\psi_i\rangle |\phi_i\rangle = \sum_n c_n |n\rangle |\phi_i\rangle \xrightarrow{H_\text{int}} \sum_n c_n |n\rangle |\phi_n\rangle \ . \tag{7}$$

Thus, if the quantum system initially is in a superposition of states, then linearity of quantum mechanics demands the probe , too, to be in a superposition of its states.

There is no *a priori* reason not to apply quantum mechanics, and, in particular the above treatment, to objects used as a probe that are made up of a large number of elementary constituents each of which is perfectly described by quantum mechanics. E. Schrödinger (Schrödinger 1935) illustrated how quantum theory, if applied to macroscopic objects, may lead to predictions that are not in agreement with our observations. He imagined a cat coupled to an individual quantum system that may exist in a superposition of states, say $|e\rangle$ and $|g\rangle$ . The apparatus is constructed such that if the quantum system is in $|e\rangle$ , the cat remains untouched, whereas state $|g\rangle$ means the cat will be killed by an intricate mechanism. The formal quantum mechanical description of this situation leads to the conclusion that the cat is in a superposition state of being dead and being alive, once the quantum system assumes a superposition state.

If the cat is replaced by an apparatus that is used to measure the state of the quantum system, one immediately sees that the Schrödinger's gedanken experiment illustrates part of the measurement problem in quantum mechanics: Why does a macroscopic probe correlated to the quantum system's state not exist in a superposition of its possible states, but instead always assumes one *or* the other?

The Kopenhagen interpretation solves this contradiction between quantum mechanical predictions and actual observations by postulating that quantum mechanics does not apply to a classical apparatus. Following this interpretation there exists a border beyond which quantum mechanics is no longer valid. This, of course, provokes the questions where exactly this borderline should be drawn and what parameter(s) have to be changed in order to turn a given quantum system into a classical device.

The mathematical counterpart of this view was formulated by von Neumann: he postulated two possible time evolutions in quantum mechanics (von Neumann 1932): One is the unitary time evolution that a quantum system undergoes according to Schrödinger's equation in absence of any attempt to perform a measurement (von Neumann's "zweiter Eingriff" or "second intervention"). This evolution is reversible. The other process is the irreversible quasi instantaneous time evolution when a measurement on the system is performed. It leads to a projection of the wave function on one of the eigenfunctions of the measured observable (called the "first intervention" by von Neumann.)

The theory of decoherence (Zurek 1991, Giulini et al. 1996) answers the question how a *superposition* of a quantum system in the course of a measurement is reduced to a state described by a local diagonal density matrix (after tracing out the probe degrees of freedom), a mathematical entity describing possible alternative outcomes, but not a superposition of states. We will consider this approach in more detail in the following paragraphs.

A cavity-QED experiment similar to the gedanken experiment envisioned by Schrödinger is realized by first preparing a Rydberg atom in a superposition of two internal energy eigenstates $|e\rangle$ and $|g\rangle$ (Brune et al. 1996). Then, this quantum system is sent through a cavity containing an electromagnetic field in a Glauber state (a coherent state corresponding to the cat in the gedanken experiment), $|\alpha\rangle$ whose phase is changed by dispersive interaction (no energy exchange takes place between atom and field) depending on the state of the atom. The combined atom-field state after the interaction reads

$$|\Psi\rangle_{\text{atom,cav}} = 1/\sqrt{2}(|e\rangle|\alpha e^{i\varphi}\rangle_C + |g\rangle|\alpha e^{-i\varphi}\rangle_C). \tag{8}$$

The decay of this coherent superposition of probe states correlated with a quantum system (Rydberg atom) towards a statistical mixture was indeed experimentally observed and quantitatively compared with theoretical predictions (Brune et al. 1996). It could be shown that the decay of the superposition becomes faster with increasing distinguishability of the two probe states involved in the measurement of the quantum system.

This decay from a superposition towards a statistical mixture is monitored by sending a second atom through the cavity (a time $\tau$ after the first atom) and detecting this second atom's state after it has interacted dispersively with the cavity field. The analysis of the correlations between the first and second atom's measurement results then reveals to what degree the off-diagonal elements of the density matrix (the coherences, created through the interaction with the first atom) describing the cavity field have decayed at time $\tau$ when the second atom was passing through the cavity (Maitre et al. 1997).

(Gedanken) experiments on quantum complementary, too, have dealt with the influence of correlations and measurements on an observed system. As an example we consider first the diffraction of electrons when passing through a double slit resulting in an interference pattern on a screen mounted behind the double slit (Feynman et al. 1965, Messiah 1976). Any attempt to determine the path the electrons have taken, that is through which slit they passed, destroys the interference pattern. This can be explained by showing that the act of position measurement imposes an uncontrollable momentum kick on the electrons in accordance with Heisenberg's uncertainty principle ((Bohr 1949), reprinted in (Bohr 1983).) This is to be regarded as a local physical interaction (Knight 1998).

In (Scully et al. 1991) it is shown by means of a gedanken experiment, without making use of the uncertainty principle, that the loss of interference may be caused by a nonlocal correlation of a welcher weg detector with the observed system: An atomic beam is detected on a screen after it has passed through a double slit. After having passed the double slit, the wave function describing the center-of-mass (COM) motion of the atoms is

$$\Psi(\vec{r}) = \frac{1}{\sqrt{2}} \left( \psi_1(\vec{r}) + \psi_2(\vec{r}) \right) \tag{9}$$

where the subscripts 1 and 2 refer to the two slits. The probability to detect an atom at location $\vec{R}$ on the screen is then given by

$$\left| \Psi(\vec{R}) \right|^2 = \frac{1}{2} |\psi_1(\vec{R})|^2 + |\psi_2(\vec{R})|^2 + \psi_1^* \psi_2 + \psi_2^* \psi_1 \tag{10}$$

where the last two terms are responsible for the appearance of interference fringes on the screen.

Now an empty (vacuum state) micromaser cavity is placed in front of each slit and the atoms are brought into an excited internal state, $|e\rangle$ before they reach one of the cavities. The interaction between atom and cavity is adjusted such that upon passing through a cavity an atom will emit a photon in the cavity and return to its lower state, $|g\rangle$. Consequently, the combined state of atomic COM wave function and cavity field is now an entangled one and reads

$$\Psi(\vec{r}) = \frac{1}{\sqrt{2}} \left( \psi_1(\vec{r}) |1\rangle_1 |0\rangle_2 + \psi_2(\vec{r}) |0\rangle_1 |1\rangle_2 \right) . \tag{11}$$

Here, the state ket representing a cavity field is labelled with the number of photons present in the cavity, and the subscripts indicate in front of which slit the respective cavity is placed. Calculating again the probability distribution on the screen now gives

$$\left| \Psi(\vec{R}) \right|^2 = \frac{1}{2} |\psi_1(\vec{R})|^2 + |\psi_2(\vec{R})|^2 + \psi_1^* \psi_2 \langle 1|0\rangle_1 \langle 0|1\rangle_2 + \psi_2^* \psi_1 \langle 0|1\rangle_1 \langle 1|0\rangle_2 . \tag{12}$$

The two last terms responsible for the appearance of interference fringes disappear, since the cavity states are orthogonal, and with them the interference pattern on the screen. It is emphasized in (Scully et al. 1991) that the welcher weg detector functions without recoil on the atoms and negligible change of the spatial wave function of the atoms. The atoms, after having interacted with the welcher weg detector behave like a statistical ensemble, and the loss of the atomic spatial coherences is due to the nonlocal correlation of the atom with the detector. Such a correlation is generally produced in every welcher weg scheme, but its effect of suppressing interference is often covered by local physical back action on the observed quantum object (Dürr, Nonn & G.Rempe 1998).

The ability of the atomic COM wave function to display interference can be restored in this gedanken experiment by erasing the welcher weg information ((Scully et al. 1991) and references therein, (Scully & Walther 1998).) However, the interference is regained only, if the detection events due to atoms arriving at the screen are sorted according to the final state of the device used to erase the Welcher Weg information (a detector for the photons in this gedanken experiment). Experiments along these lines have demonstrated such quantum erasers (Kwiat et al. 1992, Herzog et al. 1995, Chapman et al. 1995).

Here, we have considered the extreme case that complete information on the atoms path is available and the interference disappears completely. A general quantitative relation between the amount of welcher weg information stored in a detector and the visibility of interference fringes has been given in (Englert 1996). In order to verify this relation, a welcher weg experiment using an atom interferometer was carried out and is described in (Dürr, Nonn & Rempe 1998$b$, Dürr, Nonn & Rempe 1998$a$). In that experiment the amount of information stored in the detector and the contrast of interference fringes were determined independently.

The first part of the cavity-QED experiment described in (Brune et al. 1996) and outlined above (that is, before the second atom is send through the cavity) can be interpreted as an atom interferometer with a welcher weg detector in one of the arms of the interferometer (compare also (Gerry 1996)): Before the atom enters the cavity a coherent superposition of its energy eigenstates $1/\sqrt{2}(|e\rangle + |g\rangle)$ is prepared by applying a $\pi/2$-pulse to the atom. The analogy with an optical Mach-Zehnder interferometer where a photon is sent along one of two possible paths after the first beam splitter (in a classical view) is manifest in the fact that the atom may cross the cavity either in state $|e\rangle$ or state $|g\rangle$ (again classically speaking). After the atom has passed through the cavity, a second $\pi/2$-pulse is applied corresponding to the second beam splitter (or combiner) in an optical interferometer.

Placing a photo detector in one of the arms of the Mach-Zehnder interferometer would reveal information on which path the photon took. Here, the coherent field in the cavity that undergoes a phase shift correlated to the atom's state acts as a welcher weg detector. The cavity field does not act as a "digital" detector indicating the state of the atom with certainty. Instead, the two coherent field components correlated with the two atomic states may have some overlap (i.e., $\langle \alpha e^{i\varphi} | \alpha e^{-i\varphi} \rangle \neq 0$) such that they cannot be distinguished with certainty. Consequently, the correct state of the atom could only be inferred with probability below unity, if a measurement of the cavity field were performed. Therefore, the interference fringes do not completely disappear, but instead a reduced contrast of the fringes is observed (Fig. 3 in (Brune et al. 1996).)

After the welcher weg detector (the field in the cavity-QED experiment) and the atom have become entangled, the atom's capability to display interference vanishes. If only the atom is considered, that is, only one part of the entangled entities quantum system and quantum probe, then it appears as if the atom had been reduced to a statistical mixture of states as opposed to a coherent superposition. This is evident when considering the reduced density matrix of the atom obtained by "tracing out" the probe degrees of freedom. By applying a suitable global operation on probe and the atom together, the capability of the atomic states to show interference can be restored. This has been demonstrated in a different experiment where the Welcher Weg information is encoded in the photon number instead of the phase of the field (Bertet et al. 2001). Thus, the reversibility of the interaction of system and probe is demonstrated.

The quantum probe itself – in the experiment described in (Brune et al. 1996) represented by the mesoscopic cavity field initially prepared in a superposition of two states by the interaction with the atom – eventually undergoes decoherence: photons escaping from the resonator into the environment lead to entanglement between the atom, cavity field, and the previously empty, but now occupied "free space" modes of the electromagnetic field. Finally, this process results in a local diagonal density matrix (after tracing out the "free" field modes) describing a statistical mixture of the state of the atom (system) *and* the cavity field (probe). That is, the outcome of any subsequent manipulation of only the atom and/or cavity field will be characterized by the initial absence (before this further manipulation takes place) of coherent superpositions.

This argument can, of course, be extended further, including into the description also the environment with which the photons escaping from the cavity may eventually interact. Taking this argument consecutively further, always leaves behind some entities (the atom, cavity field, "free" field, ...) that will behave as statistical mixtures, if the next entity is not included in the theoretical description and further experiments. In practical experiments it seems impossible to include the whole chain of entities in further manipulations. Therefore, for all practical purposes, the correlation established between system, probe, and environment irreversibly destroys the system's and probe's superposition state. For a macroscopic environment (e.g., a measurement apparatus) this reduction to a statistical mixture occurs quasi-instantaneous (Giulini et al. 1996).

In the considerations to follow, we divide the measurement apparatus, used to extract information about the state of a quantum system, into a quantum probe that interacts with the observed quantum system and a macroscopic device (called "apparatus" henceforth) coupled to the probe and yielding macroscopically distinct read-outs. The measurement process is then formally composed of two stages (as outlined above; (von Neumann 1932, Alter & Yamamoto 2001, Braginsky & Khalili 1992)). First a unitary interaction between quantum probe and the quantum system takes place. Then the quantum probe is coupled to the apparatus that indicates the state of the probe by assuming macroscopically distinct states (e.g., pointer positions.) The "environment" in the cavity-QED experiment described above takes on only part of the role of the apparatus: in principle, information about the probe's state is available in the environment after a time determined by the decay constant of the cavity field. However, it will be difficult for an experimenter to extract this information by translating it into distinct read-outs of a macroscopic meter.

## 3.2 Measurements on individual quantum systems

The theory of decoherence explains the appearance of local alternatives with certain statistical weights instead of coherent superpositions in quantum mechanical measurements. But it does not give an indication of which eigenstate the probe (and consequently the system) will be reduced to as the final result of the measurement. The density matrix describing system and probe, according to decoherence theory, becomes diagonal as a result of the interaction with the apparatus, but, in general, still has more than one diagonal element larger than zero. This will be a valid description, if after a measurement has been performed on an *ensemble* of quantum systems, further manipulations of this ensemble are carried out. However, such a density matrix is not in agreement with the experimental observation that after a measurement has taken place on an *individual* quantum system, and a particular eigenvalue of the measured observable has been obtained, subsequent measurements again yield the same result. After such a measurement, the state of this single quantum system has to be described by the density operator $\rho = |n\rangle\langle n|$ of a pure state, that is all diagonal elements vanish except one. Decoherence cannot explain or predict what particular outcome a given measurement on an individual quantum system has (i.e., which diagonal element becomes unity.) The measurement of a single quantum system corresponds to a projection of the system's (and probe's) wave function on a particular eigenstate $|n\rangle$ ($|\phi_n\rangle$) in accord with von Neumann's first intervention (the projection postulate.)

According to the projection postulate, the wave function of the object collapses into an eigenfunction of the measured observable due to the interaction between the measurement apparatus and the measured quantum object. The result of the measurement will be the corresponding eigenvalue. One could suspect that the statistical character of the measurement process described by the projection postulate is due to incomplete knowledge of the quantum state of the measurement apparatus. However, von Neumann showed that the measurement process remains stochastic even if the state of the measurement apparatus were known (chapter VI.3 in (von Neumann 1932).) We have seen that decoherence can account for the quasi-instantaneous disappearance of superpositions and the appearance of distinct measurement outcomes with certain probabilities, but not for the "choice" of a particular outcome of a measurement on an individual system (the projection postulate, too, does not explain this last point.)

A quantum mechanical wave function can be determined experimentally from an ensemble experiment: Either a series of measurements is performed on identically prepared single systems, or a single measurement measurement on an ensemble of identical systems is carried out (von Neumann 1932, Alter & Yamamoto 2001, Raymer 1997). The wave function is interpreted as a probability amplitude that defines a probability density $P(a) = |\langle\psi|\psi\rangle|^2$, the distribution of possible results $a \in \mathbb{R}$ of measurements of an observable $\hat{A}$. The corresponding expectation value $\langle a \rangle = \langle\psi|\hat{A}|\psi\rangle$ defines the center position and the width $\langle(\triangle a)^2\rangle = \langle a^2\rangle - \langle a\rangle^2$ of the probability density. It is possible to determine *both* quantities in an *ensemble* measurement, and therefore to infer the quantum wave function up to a phase.

The expectation value might be estimated from

$$\langle a \rangle \approx \frac{1}{N} \sum_{n=1}^{N} a_n \qquad (13)$$

where the $a_n$ are the results of local measurements on identically prepared independent quantum systems. In a similar way $\langle a^2 \rangle$ and $\langle a \rangle^2$ can be determined, and thus $\langle (\Delta a)^2 \rangle$. Though it is possible to determine an unknown quantum wave function from an ensemble measurement, it is impossible with a *single* quantum system, neither from a single measurement nor from a series of subsequent measurements. If $a_1$ is the result of a *single* measurement, the estimated expectation value $\langle a_s \rangle = a_1$, and in general $\langle a \rangle \neq \langle a_s \rangle$. The quantum uncertainty, $\langle (\Delta a)^2 \rangle$ of the measured observable remains undetermined, since $\langle a_s^2 \rangle = \langle a_s \rangle^2 = a_1^2$. Even if a *series* of measurements on the *same* single system is performed, it is not possible to infer the probability distribution $P(a)$. The results of $N$ subsequent measurements of $\hat{A}$ on a single system are not independent, and one will obtain the same eigenvalue $a_1 = a_2 = ... = a_N$ for every observation. The estimated expectation value will then again be $\langle a_s \rangle = \langle a_1 \rangle = a_1$, and the variance $\langle \Delta a \rangle = 0$ (compare also chapter 2 in (Alter & Yamamoto 2001).)

In this sense, quantum mechanics is *not* an ergodic theory, in contrast to classical statistics where a series of measurements on a single system is equivalent to a single measurement on an ensemble of identical systems. Only if an observed single quantum system is identically prepared in advance of every subsequent measurement, a series of measurements on a single quantum system is equivalent to a single measurement of an ensemble of quantum systems.

In general, it is not possible to predict the outcome of a measurement on an individual quantum system with certainty, even if complete knowledge of the initial quantum wave function is available. The obtained results are statistical, if the system is not initially prepared in an eigenstate of the observable being measured. On the other hand, if the initial state is an eigenstate, then the measurement is compatible to the preparation. Therefore, even a single measurement may yield *partial* information about the systems initial state: If an eigenvalue is obtained corresponding to a particular eigenstate $|n\rangle$, the observed system was initially not in an eigenstate $|n'\rangle$ orthogonal to the measured one.

So far, in the discussion of measurements on quantum systems we have not explicitly considered the case of negative result measurements (for a recent review see (Whitaker 2000).) We will restrict the following discussion to quantum mechanical two-state systems for clarity. In some experimental situations (real or gedanken) the apparatus coupled to the quantum probe and quantum system, may respond (for example by a "click" or the deflection of a pointer) indicating one state of the measured system, or not respond at all indicating the other. Such measurements where the experimental result is the absence of a physical event rather than the occurrence of an event have been described, for instance, in (Renninger 1960, Dicke 1981). A negative-result measurement or observation leads to a collapse of the wave function *without* local physical interaction involved between measurement apparatus and observed quantum system. This will be discussed in more detail in the following paragraphs. In particular, the meaning of the concept "local physical interaction" is looked at in this context.

The situation described above is analogue to a gedanken experiment depicted in chapter 3.3.2.3 in (Giulini et al. 1996). There, a Stern-Gerlach (SG) apparatus is considered that is oriented to yield at its exit spin-1/2 particles with their spin pointing either in the positive or negative $z$ direction. Particles with different spin directions propagate along spatially separate trajectories upon exiting the SG device (Fig. 4). A non-absorbing detector, $D_A$ is placed in only one of the exit "channels" A of the device SG such that it will register a particle passing through. If the particle takes the other channel B (corresponding to the orthogonal spin direction) then the detector $D_A$ does not respond. If for times greater than $t_A$ (the time needed for the particle to travel from the entrance of SG to $D_A$,), $D_A$ has not indicated the passing through of a particle, and if an additional auxiliary detector $D_B$ were used in channel B, placed far away from SG and $D_A$ (we take $t_B \gg t_A$ ), then this detector $D_B$ would register the particle with certainty at time $t_B$ after it was launched at the entrance of SG. Even though in this example, in a classical sense, the particle and the detector $D_A$ never interacted, (since they are located in regions of space separated by a distance much larger than the deBroglie wavelength of the particle), the mere possibility for detection may change the behavior of the particle: a spin-1/2 at the entrance of SG initially prepared in a *superposition* of eigenstates of $\sigma_z$ is effectively reduced to an eigenstate of $\sigma_z$. In (Giulini et al. 1996) it is argued: "The claim that the particle did not interact at all with the detector $[D_A]$ in the case of a spin-down result [detector $D_A$ does not "click"] must be wrong, since a [superposition state] is different from an ensemble of $z-$up and $z-$down states." In this argument, the change of a quantum state is taken as a sufficient condition for "interaction" between the quantum system and some device (quantum or macroscopic). What is termed "interaction" in (Giulini et al. 1996) we consider as the consequence of a negative result measurement, a measurement not involving a local physical interaction between detector and system.

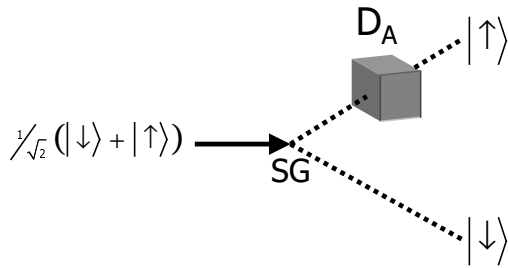

Figure 4: Illustration of a negative result measurement using a Stern-Gerlach apparatus (SG) with a nonabsorbing detector ($D_A$) at one of its exits (see text).

If detector $D_A$ did not respond (a negative-result measurement occurred,) then the quantum state has nevertheless changed as described above: a coherent superposition is reduced not only to a statistical mixture, but to a definite state. In a classical sense, no interaction between the particle and the detector took place, since the particle is travelling along path B for $t > t_A$. This we consider the absence of *local* physical interaction. The Hamiltonian describing the quantum system (spin-1/2 after having passed through SG) and the detector ($D_A$) contains a term coupling the spin system to the detector $D_A$, *only* if the spin is in the $z-$up state, thus describing a conditional local physical interaction (which is absent if the spin is in $z-$down state.) In general, a Hamiltonian determines eigenstates and -values, and fixes a range of *possible* measurement outcomes, some of which may be obtained without local physical interaction.

# 4 Impeded quantum evolution: the quantum Zeno effect

The non-local character of negative result measurements manifests itself in an effect that Misra and Sudarshan named "Zeno's paradox in quantum theory" (or "quantum Zeno effect") alluding to the paradoxes of the greek philosopher Zeno of Elea (born around 490 BC), who claimed that motion of classical objects is an illusion. Zeno illustrated his point of view with various examples one of which is the following: before an object can reach a point at a distance $d$ from its present location, it must have passed through the point at distance $d/2$. Carrying this argument further, it means that infinitely many points have to be passed in finite time before $d$ can be reached. Therefore motion is not possible according to this argument. Modern Mathematics resolves this apparent "paradox" making use of real numbers and convergent infinite series. In the quantum domain the notion "quantum Zeno effect" refers to the impediment or even suppression of the dynamical evolution of a quantum system by frequent measurements of the system's state (see, for instance, (Khalfin 1968, Fonda et al. 1973, Misra & Sudarshan 1977, Beige & Hegerfeldt 1996, Home & Whitaker 1997) and references therein, and also chapter V.2 in (von Neumann 1932).)

How does a quantum system behave, whose evolution in time is unitary, under repeated measurements separated by the time $\triangle t$? This will be considered for the case of ideal measurements, that is, the measurement is instantaneous and leaves the quantum system in an eigenstate of the observable being measured (Beige & Hegerfeldt 1997). Let $|a\rangle$ be an eigenstate of observable $\hat{A}$, and $\hat{P}_a = |a\rangle\langle a|$ the corresponding projector. If a quantum system, initially prepared in state $|\psi(0)\rangle$, undergoes ideal measurements at times $t_1, t_2, ....$ with $\triangle t = t_i - t_{i-1}$ ($i = 1, 2, \ldots, N$), then after $N$ successive measurements the system is found in state

$$|\psi(t_N, 0)\rangle = \hat{P}_a \hat{U}(t_N, t_{N-1}) \hat{P}_a \ldots \hat{P}_a \hat{U}(t_1, 0)|\psi(0)\rangle \tag{14}$$

where $\hat{U}(t_i, t_{i-1})$ denotes the unitary time evolution operator for the quantum system between two measurements. The probability to find the quantum system in the state $|a\rangle$ after $N$ ideal measurements is

$$P_a = |\langle a|\psi(t_n, 0)\rangle|^2 \tag{15}$$

$$= |\langle a|\hat{U}(t_1, 0)|\psi(0)\rangle|^2 \prod_{i=2}^{n} |\langle a|\hat{U}(t_i, t_{i-1})|a\rangle|^2 , \tag{16}$$

and an expansion gives:

$$|\langle a|\hat{U}(t_i, t_{i-1})|a\rangle|^2 \simeq 1 - \triangle t^2 (\triangle \hat{H})^2 \tag{17}$$

If the time interval between subsequent measurements goes to zero, $\triangle t \to 0$, then $P_a$ tends to 1 and the survival probability becomes $P_a = |\langle a|\psi(0)\rangle|^2$. This simple argument shows that a quantum system initially in a state $|\psi_i\rangle$ turn into an eigenstate $|a\rangle$ under repeated ideal measurements for $\triangle t \to 0$ with probability $|\langle a|\psi(0)\rangle|^2$ (von Neumann 1932). If $|\psi(0)\rangle = |a\rangle$, the system remains in $|a\rangle$ for $\triangle t \to 0$.

For short times the survival probability of the state will be proportional to $t^2$, and the decay rate of this state is proportional to $t$. In contrast, an exponential decay occurs with a constant rate, and a decay for time $t_1$, followed by an interruption (or measurement), followed by further decay for a time $t_2$ is equivalent to uninterrupted decay for a time $t_1 + t_2$ (Home & Whitaker 1997).

A simple quantum system to demonstrate the quantum Zeno effect is a stable two-level system (states $|0\rangle$ and $|1\rangle$ with energy separation $\hbar\omega_0$) driven by a resonant harmonic perturbation. After unitary time evolution of duration $\triangle t$, the probability of finding the system in the initially prepared eigenstate, e.g. $|0\rangle$ (the survival probability,) $P_0 = \cos^2(\theta/2)$, where $\theta = \Omega \cdot \triangle t$, and $\Omega$ is the Rabi frequency. The corresponding transition probability $P_1 = \sin^2(\theta/2)$. For small time intervals $\Delta t$ the survival probability becomes

$$P_0 = \cos^2\left(\frac{\Omega \cdot \Delta t}{2}\right) \simeq 1 - \frac{\Omega^2 \Delta t^2}{4} \tag{18}$$

displaying the initial quadratic time dependence required for the quantum Zeno effect.

When an ideal measurement is carried out at the end of a period of evolution $\Delta t$, the quantum system is reset to one of its eigenstates. If during time evolution one performs $q$ successive ideal measurements a time $\Delta t$ apart, the survival probability to find the system in the initial eigenstate in measurement $q$ under the condition that it was found $q - 1$ times in this state before,

$$P_{00} = \cos^{2\cdot q}(\theta/2) \ . \tag{19}$$

In Figure 5a) $P_{00}$ is shown for several values of $\theta$. On the other hand, if no measurements were performed and the system evolved coherently, the (*a priori*) probability that the system is in the initial state after time $t = n \cdot \Delta t$ is $P_{\text{coh}}(n) = \cos^2(n \cdot \theta/2)$ (Figure 5b).

In their original proposal of the quantum Zeno effect Misra and Sudarhan used the term "quantum Zeno paradox" for the case of "freezing" the system to a particular state by means of continuous observation of the systems unitary evolution (Misra & Sudarshan 1977), while the term "quantum Zeno effect" was used to characterize its impediment (Peres 1980, Pascazio & Namiki 1994, Cook 1988). Other authors distinguish between unitary evolution of the quantum system and exponential decay of an unstable system, and the suppression of the latter is regarded paradoxical (Block & Berman 1991). In (Home & Whitaker 1997) it is pointed out that the the quantum Zeno effect is a quantum effect due to the initial quadratic time dependence of quantum mechanical evolution. In contrast, a strictly exponential decay is a classical concept. In order for the quantum Zeno effect to take place when the system is characterized by an exponential decay, deviations from the exponential law at short times would be required (an initial quadratic time dependence.) For unstable *quantum* systems these short time deviations were indeed predicted (Winter 1961, Fonda et al. 1978), and observed experimentally in the tunnelling of atoms from a trapped state into the continuum (Wilkinson et al. 1997). The use of the term quantum

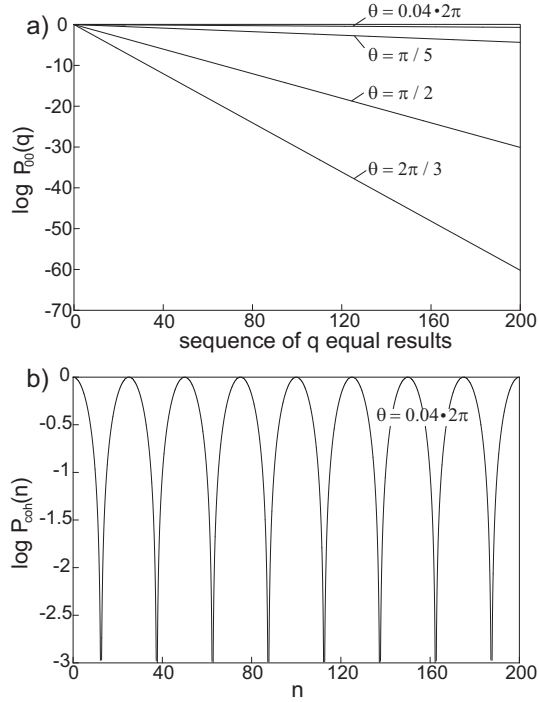

Figure 5: a) Probability, $P_{00}(q)$ to find a harmonically driven two-state system in the initially prepared state $|0\rangle$ in each one of $q$ successive measurements, resulting in uninterrupted sequences of $q$ equal results (for several nutation angles $\theta = \Omega \Delta t$.) b) *A priori* probability for the system to be in state $|0\rangle$ after time $n \cdot \Delta t$, if no measurements are performed.

Zeno paradox to describe the inhibition of an exponential decay, therefore, seems inappropriate, since it requires the same initial time dependence to take place as in the unitary case.

What can be regarded as paradoxical about the quantum Zeno effect? In a comprehensive review by (Home & Whitaker 1997), it is stressed that the paradoxical aspect is the retardation of evolution without any back action on the observed quantum system during the measurement process, as a consequence of negative result measurements. In the terminology used in this article this would correspond to the absence of local physical interaction in the course of a negative result measurement. The mere presence of the macroscopic measurement apparatus (like the detector $D_A$ in the Stern-Gerlach scheme discussed above) may affect the quantum system due to the nonlocal correlation between the two. (Home & Whitaker 1997) suggest that a nonlocal negative result measurement on a microscopic system characterizes the quantum Zeno *paradox*.

It seems sensible to extend this definition of the quantum Zeno paradox to two more classes of measurements that are *not* of the negative-result type (Toschek & Wunderlich 2001): i) measurements free of back action (quantum nondemolition measurement (Braginsky & Khalili 1992, Alter & Yamamoto 2001)), that in fact give rise to positive results, and ii), measurements whose back action cannot account for the retarding effect. In both cases the local interaction (in connection with positive results) alone, cannot explain the change in the dynamics of the quantum system, and experiments that obey those criteria would show the quantum Zeno paradox.

In the theoretical considerations at the beginning of this section state vectors have been used, that is, the behavior of individual quantum systems was investigated. Why is it necessary to carry out experiments on the quantum Zeno paradox with individual quantum systems? Important work related to this question is found, for instance, in (Spiller 1994, Alter & Yamamoto 1997, Nakazato et al. 1996, R.Wawer et al. 1998). The next paragraphs will be concerned with some aspects connected to this question. A more detailed discussion, concerning in particular experiments with trapped ions, is given in (Toschek & Wunderlich 2001, Wunderlich et al. 2001).

The original formulation of the quantum Zeno effect considered the probability for the observed system to stay in its initial state throughout the time interval during which measurements are made. It has been pointed out (Nakazato et al. 1996) that in ensemble measurements it is not possible to record this probability, unless different subensembles are chosen for each measurement, conditioned on previous measurement results. In usual ensemble experiments only the *net* probability of making or not making a transition from 0 to 1 *after* a series of $N$ measurements is recorded and calculated to interpret the experiment. Experiments with single quantum system permit to record each individual measurement result and thus to select sequences of results where the system remained in its initial state.

Furthermore, by making a series of measurements on an ensemble of identically prepared quantum systems the effect of the measurement on the quantum systems' evolution cannot be distinguished from mere dephasing of the members of the ensemble (Spiller 1994). (For example, collisions between atoms lead to dephasing of the atoms' wave functions.) Both processes lead to the destruction of coherences (off-diagonal elements of the density matrix) and give rise to identical dynamical behavior when the quantum system, after the measurement has been performed or dephasing has set in, will be subjected to subsequent manipulations. When investigating the quantum Zeno paradox we are interested in the change in the system's dynamics conditioned on the outcome of the measurement, in particular of negative-result measurements. Since dephasing of an ensemble as described above might occur *independently* of the measurement results, the question whether and how a series of particular measurement results is correlated with, and influences the quantum system's dynamics cannot be answered by an ensemble experiment. One might argue that dephasing *is* a measurement no matter how it comes about. During the process where the wave functions of the members of an otherwise isolated ensemble loose their initial phase relation via some mutual interaction (they have been identically prepared initially) correlations are established between members of this ensemble. This, however, does not establish a measurement of the initial state of the quantum systems.

In accordance with the discussion in section 3, the following condition is taken as a necessary one to constitute a measurement: some correlation is established between the quantum system (or an ensemble of quantum systems) and the "outside world" (not described by the elements of the Hilbert space(s) of the quantum system(s) under investigation.) This could be an apparatus that assumes classically distinct states correlated to the quantum system's state.

## 4.1 Experiments

An experiment with several thousand Be$^+$ ions stored in an electromagnetic trap (Itano et al. 1990, Itano et al. 1991) (based on a proposal by Cook for a single ion (Cook 1988)) indeed shows the reduction of the transition probability between coherently driven hyperfine states (here, we label them $|0\rangle$ and $|1\rangle$) when the ions' state was frequently probed. Probing the ions' state is achieved by irradiating them with light resonantly coupling one of the hyperfine states to a third level $|2\rangle$ such that scattering of light occurs, if and only if an ion occupies, say state $|0\rangle$.

After initial preparation of the ions in $|0\rangle$, they are driven by a microwave $\pi$-pulse inverting the population of the hyperfine states. To investigate the effect of repeated measurements on the transition probability between states $|0\rangle$ and $|1\rangle$, the sample of ions is irradiated, during the driving pulse, by $N$ resonant probe light pulses. At the end of the microwave pulse the population of state $|0\rangle$ is measured by again applying a probe pulse and detecting scattered light. The outcome of the experiment shows a reduction of the observed transition probability in agreement with the predicted *net* transition probability

$$P_{e1}(T) = \frac{1}{2}[1 - \cos^N(\theta/N)] \qquad (20)$$

where $\theta = \pi$, and $T$ is the duration of the microwave pulse. The index $e1$ indicates that the ions in this ensemble experiment are found in state $|1\rangle$ irrespective of the results of intermediate probing (taking place between initial preparation and final probing $N$). The corresponding survival probability $P_{e0} = 1 - P_{e1}$. The theoretical transition probability is derived from a quantum mechanical model taking into account the probe light pulses that leave the population of states $|0\rangle$ and $|1\rangle$ unchanged and just set the coherences to zero (Itano et al. 1990).

The inhibition of the quantum system's evolution was considered to be a consequence of measurements (light scattering) frequently projecting the ions back to their initial state. In (Frerichs & Schenzle 1991), calculations of the dynamics of such a three level system are reported. It is deemed not necessary to invoke the notion of measurement together with state reduction to explain that the quantum system's evolution was impeded in the experiment. Instead, the retardation of the 2-state system's evolution is interpreted as a dynamical effect that can be explained when the third level is included in the quantum mechanical description (Frerichs & Schenzle 1991, Block & Berman 1991, Gagen & Milburn 1993). Indeed, good agreement is found with experimental data gathered from the ensemble of Be$^+$ ions. This is not so surprising, since the results of the experiment are expectation values of an ensemble of ions, and one would not expect quantum mechanics to fail in predicting the correct ensemble average. Each measurement leads to a diagonal density matrix describing the ions ($\rho_{00} \neq 0 \neq \rho_{11}$), however, with *both* diagonal elements different from zero. However, the paradoxical aspect of quantum mechanics, and in particular of quantum Zeno, comes into focus when the eigenvalue of every single system as a result of a measurement is revealed.

Both state reduction and Bloch equations may lead to identical results when measurements on an ensemble are performed. This has been shown in (Power & Knight 1996) and (Beige & Hegerfeldt 1996) where the ensemble quantum Zeno experiment with Be$^+$ ions is simulated using quantum jump techniques in order to test whether the projection postulate is applicable to describe the observed results. It is pointed out that for an ensemble, the quantum trajectories produced by the quantum jump approach reproduce the density matrix probabilities resulting from the Bloch equations. In the latter model the decay of the coherences is due to coupling of the driven transition to the strong monitor transition (Frerichs & Schenzle 1991, Power & Knight 1996). Therefore, to understand the ensemble averaged relaxation, it is not necessary to refer to state reduction. In (Beige & Hegerfeldt 1996) it is suggested that under particular conditions (that were fulfilled in the experiment) the projection postulate is a useful tool that gives the right results. On the other hand, in (Power & Knight 1996) it is pointed out that the Bloch equations do not hold for the description of the quantum Zeno effect with a *single* ion, since one of the diagonal elements of the density matrix disappears whereas in an ensemble, in general, both diagonal elements assume nonzero values.

Another aspect (connected to the above argument) to mention is that in the Be$^+$ experiment only the net transition probability at the end of the microwave pulse is recorded. Intermediate back-and-forth transitions between states $|0\rangle$ and $|1\rangle$ of individual members of the ensemble, as well as correlated transitions of ions, could not be detected. In (Nakazato et al. 1996) it is worked out that, if one takes into account the result of every intermediate measurement, the probability in equation 20 describes *not* the quantum Zeno effect of a two-level system, equation 20 includes these intermediate back-and-forth transitions, which means the system does not necessarily stay in the initial state. The correct description is the one in equation 19. Both equation 20 and 19 imply that an ensemble (for $N \to \infty$ nonselective measurements) and a single quantum system ($q \to \infty$ selective measurements) are found in the initial state. However, for small $N$ ($q$), expression 20 and 19 yield markedly different results (section 4.3.)

The experiment described in (Kwiat et al. 1995) aimed at the demonstration of an optical version of the quantum Zeno effect. Based on a suggestion put forth in (Elitzur & Vaidman 1993), the propagation of a photon in a sequence of Mach-Zehnder interferometers is restricted to only one arm of the interferometers due to interaction-free measurements. Even though the outcome of the experiment obeys the mathematics of the quantum Zeno effect, the physics seems different as pointed out in (Home & Whitaker 1997) where it is argued that the result of this experiment is explicable, as far as the quantum Zeno effect is concerned, in terms of classical physics. A modification of this experiment shows the polarization rotation of photons to be impeded because of an interaction-free measurement within the Mach-Zehnder interferometer (Kwiat et al. 1999). As in the previous experiment the mathematics of the quantum Zeno effect describes well the dynamical behavior of the system. According to the arguments in (Home & Whitaker 1997, Whitaker 2000), it appears that again the physics necessary for the quantum Zeno effect is not involved. An experiment that can be classically described gives equivalent results: the rotation of the polarization of light passing through an optically active substance is retarded by means of a sequence of polarization analyzers (Peres 1980).

Recently an experiment was performed to demonstrate the quantum Zeno effect and the anti Zeno effect in an unstable system (Fischer et al. 2001). The anti Zeno effect describes an acceleration of the decay of an unstable system under repeated observation (Kofman & Kurizki 2000, Facchi et al. 2001). As stated previously, the quantum Zeno effect may occur, if the short time evolution of the decay deviates from a purely exponential one (Winter 1961, Fonda et al. 1978, Wilkinson et al. 1997). Reference (Fischer et al. 2001) describes the decay via tunnelling of an ensemble of atoms trapped in an optical potential created by a standing light wave. Acceleration of the standing wave leads to a deformed potential, thus admitting tunnelling of some atoms out of the optical potential wells. The tunnelling probability shows a marked deviation from exponential decay for short times that has its origin in the initial reversibility of the decay process. Tunnelling is initiated by applying high acceleration to the atoms trapped in the standing wave for a time $t_{\text{tunnel}}$, and interrupted for time $t_{\text{interr}}$ during which the acceleration was low. The interruption of tunnelling is considered a measurement of the number of atoms that remain trapped, since $t_{\text{interr}}$ is chosen such that the fraction of trapped atoms separate in momentum space from the atoms that have tunnelled during $t_{\text{tunnel}}$. The insertion of periods of low acceleration indeed leads to a slower decay of the survival probability of trapped atoms. It seems that this experiment does not satisfy the criteria for the quantum Zeno paradox for similar reasons as the experiment with an ensemble of Be$^+$ ions outlined above. The final measurement of the spatial distribution of all atoms yields an ensemble average in agreement with the unitary time evolution predicted by the Schrödinger equation. The intermediate measurement results (obtained after periods of low acceleration) were not recorded; even if this had been the case, back-and-forth transitions between trapped and free states of individual members of the ensemble during the initial period (reversible dynamics) would have gone unnoticed.

The discussed experiments appear not suitable to demonstrate the Quantum-Zeno-Effect, or rather the quantum Zeno paradox, for they do not address a key point that makes up the nature of the effect: the retardation of the evolution of a quantum system due to a (possibly nonlocal) correlation between the observed individual quantum system and the macroscopic measurement apparatus during the repeated measurement process. This correlation leads to an irreversible change in the system's wave function and is evident even in negative result measurements where its effect is not concealed by local physical interaction. The latter, too, may indeed affect the system's transition probability under the condition of an initial quadratic time dependence. However, such a change in the time evolution is necessary but not sufficient for the quantum Zeno effect.

## 4.2 Quantum Zeno experiment on an optical transition

An experiment with a single $^{172}$Yb$^+$ ion demonstrating the quantum Zeno effect will be outlined in what follows (Balzer et al. 2000). The electronic states $S_{1/2} \equiv |0\rangle$ and $D_{5/2} \equiv |1\rangle$, connected via an optical electric quadrupole transition close to 411nm, serve as a two-level quantum system. State $|0\rangle$ is probed by coupling it to state $P_{1/2}$ via a strong dipole transition and detecting resonance fluorescence close to 369nm. The quadrupole transition $|0\rangle - |1\rangle$ was coherently driven using light emitted by a diode laser with emission bandwidth 30Hz (in 2ms). To demonstrate the retardation of quantum evolution, driving light pulses close to 411 nm alternated with probe pulses at 396nm. The duration, $\Delta t$ and the Rabi-frequency, $\Omega$ of the driving pulse were set to fixed values, and the frequency of the light field was slightly detuned from exact resonance in order to vary the effective nutation angle $\theta_{\text{eff}} = \sqrt{\Omega^2 + \delta^2} \cdot \Delta t$. The intensity and the duration of the probe field were adjusted such that the observation of resonance fluorescence results in state reduction to state $|0\rangle$, while the absence of fluorescence results in state $|1\rangle$ with near unity probability. *Each* outcome of probing was registered, and a complete record of the evolution of the single quantum system was acquired. Thus, a trajectory of "on" results (resonance fluorescence was observed) and "off" (no fluorescence, i.e. negative) results is obtained. The statistical distribution of uninterrupted sequences of $q$ equal results was found in good agreement with $P_{00}(q-1) = U(q)/U(1)$ where $U(q)$ is the normalized number of sequences with $q$ equal results, and $U(1)$ denotes the probability for this result at the beginning of the sequence. This shows the impediment of the system's evolution under repeated measurements, and thus the quantum Zeno effect. A theoretical model taking into account spontaneous decay of the $D_{5/2}$ state fits well the recorded series of "off" events (negative-result measurements) as well as to the "on" events (positive-result measurements.) It has been shown that the effect of the measurement on the ion's evolution is not intertwined with additional dephasing effects (Balzer et al. 2000, Toschek & Wunderlich 2001). The observed impediment of the driven evolution of the system's population is a consequence of the correlation between the observed quantum system and the macroscopic meter.

In this experiment the angle of nutation $\theta$ was not exactly predetermined. During the driving pulse, the system's population undergoes multiple Rabi oscillations giving an effective nutation angle $\theta_{\text{eff}} = \theta \mod 2\pi$ at the end of the interaction that varies in a small range due to not perfect experimental conditions. Therefore, the exact nutation angle was obtained from a fit of experimental data. The analysis of the experiment is further complicated by spontaneous decay from the relatively short-lived $D_{5/2}$ state (lifetime of 6ms (Fawcett & Wilson 1991)) into the $S_{1/2}$ ground state and the extremely long-lived $F_{7/2}$ state of $^{172}$Yb$^+$ (lifetime of about 10 years (Roberts et al. 1997).)In addition, the relatively short time series recorded in this experiment may cause interpretational difficulties.

## 4.3 Quantum Zeno experiment on a hyperfine transition

In this section we describe an experiment with a single $^{171}$Yb$^+$ ion whose ground-state hyperfine states are used as the quantum system to be measured. Here, the quantum Zeno paradox is demonstrated avoiding the complications associated with relaxation processes and optical pumping as in the experiment described in the previous section (Balzer, Hannemann, Reiß, Wunderlich, Neuhauser & Toschek 2002). The hyperfine transition is free of spontaneous decay and the use of microwave radiation allows for precise preparation of states with a desired nutation angle $\theta$. Sufficiently extensive data records ensure an unambiguous interpretation of these experiments.

parameters of the microwave field driving this transition are precisely defined.

A semiclassical treatment of the magnetic dipole interaction between microwave field and hyperfine states of Yb$^+$ in an interaction picture (and making the rotating wave approximation) yields the time evolution operator $U(t) =$ $\exp\left[-\frac{i}{2}t\left(\delta\sigma_z + \Omega\sigma_x\right)\right]$. (compare section 2). For $t > 0$ the ion evolves into a superposition state

$$|\psi\rangle_I = \cos\frac{\theta}{2}|0\rangle + \sin\frac{\theta}{2}e^{i\phi}|1\rangle , \qquad (21)$$

and the probability, $P_1(t)$ to find the system in $|1\rangle$ is proportional to $t^2$ for small $t$. In the experiment the resonance condition $\omega_0 = \omega$ is fulfilled to good approximation, and after time $T = \pi/\Omega$ of unperturbed evolution, a measurement of the ion's state will reveal it to be in state $|1\rangle$ with close to unit probability.

### 4.3.1 State selective detection

The relevant energy levels of the $^{171}$Yb$^+$ ion are schematically shown in Figure 1. Sufficiently long irradiating the ion with uv laser light will prepare the ion in the ground state $F = 0$ by optical pumping. The occupation of the $F = 1$ level (after interaction with the microwave field) is probed by irradiating the ion with light at 369 nm (uv laser light,) thus exciting resonance fluorescence on the electric dipole transition $S_{1/2}, F = 1 \leftrightarrow P_{1/2}$, and detecting scattered photons using a photomultiplier tube. An "on" result (scattered photons are registered) leaves the ion in state $|1\rangle$, otherwise the ion is in the $|0\rangle$ ("off" result; no photons are registered.)

While the uv light is turned on for detection of the ion's state, the ion may be viewed as a beam splitter for the incident light beam: Either the light is completely "transmitted", that is, the initially populated light mode (characterized by annihilation operator $b$) remains unchanged. This will occur with probability, $w_0$ close to unity, if the ion is in state $|0\rangle$ during the uv laser pulse (we take $w_0 = 1$ in what follows). Or, photons are scattered into some other mode $b' \neq b$ (that may be different for every scattered photon,) if the ion resides in $|1\rangle$. The latter occurs with probability $w_1$ determined by the detuning relative to the $S_{1/2}, F = 1 \leftrightarrow P_{1/2}, F = 0$ resonance, intensity, and duration of the incident uv light. For a sufficiently long uv light pulse eventually a photon will be scattered into mode $b'$, and we may take $w_1 = 1$. After one photon has been scattered into mode $b'$, the ionic state correlated with this electromagnetic field mode $b'$ is $|1\rangle$. Thus, the correlation established between the state of the light field and the ion's state is

$$\begin{aligned} |0\rangle|b\rangle &\to |0\rangle|b\rangle \\ |1\rangle|b\rangle &\to |1\rangle|b'\rangle , \end{aligned} \qquad (22)$$

and consequently

$$\alpha|0\rangle + \beta|1\rangle \to \alpha|0\rangle|b\rangle + \beta|1\rangle|b'\rangle \qquad (23)$$

In this (simplified) description $|b\rangle$ represents the em field in its initial mode, and $|b'\rangle$ stands for a different mode occupied by a single photon. Since the field states $|b\rangle$ and $|b'\rangle$ are orthogonal, the density matrix describing the ion's state (obtained by tracing over the field states) becomes diagonal, and coherences of the ion's states $|0\rangle$ and $|1\rangle$ that may have existed are no longer observable (Giulini et al. 1996). The field carries information about the ion's state, thus destroying the ion's ability to display characteristics of a superposition state in subsequent manipulations it may be subjected to.

The scattered photon in mode $|b\rangle$ may be absorbed by the photo cathode of a photo multiplier tube leading, after several amplification stages, to the ejection of a large number of photo electrons from the surface of the last dynode of the photo multiplier. This current pulse strikes the anode of the multiplier and is further amplified and finally registered as a voltage pulse by a suitable counter. Thus, the ion's state is eventually correlated irreversibly with the macroscopic environment. The irreversible correlation will actually take place much earlier in the detection chain. Irreversibility here means it is not possible, or rather very improbable, to restore the photo cathode (which would include, for instance, the power supply connected to it) to its state before an electron was ejected in response to an impinging photon.

Does the finite detection efficiency for photons (only the small fraction of about $4 \times 10^{-3}$ of scattered photons are detected during an "on" event) influence the interpretation of and conclusions drawn from the experiment described here? In order to answer this question we look in some more detail at the process of correlation between the ion's state and the 'rest of the world.'

After the first photon has been scattered from mode $b$ into an orthogonal mode $b'$, a correlation between the ion and its macroscopic environment has been established, even if this photon is not registered by the photomultiplier tube, but instead is absorbed, for instance, by the wall of the vacuum recipient housing the ion trap. Welcher weg information about the state of the ion is available, and the ion is left in a statistical mixture of states, corresponding to a density matrix with two diagonal elements different from zero (if one uses the density matrix formalism to describe an ensemble of such individual quantum systems.) The quantum Zeno experiment described below shows that the correct description of the single ion's state after a measurement pulse is either $|0\rangle$ *or* $|1\rangle$ (corresponding to a density matrix with only *one* diagonal element.) One may wonder whether (after a single photon has been scattered and absorbed by a wall) the ion is already reduced to the $F = 1$ state, or, alternatively if it is necessary for the scattered photon to hit the photo detector and thus yield a macroscopically distinct read-out for this to happen.

The second alternative does not seem plausible, since it would mean that the macroscopic photo detector plays a distinctive role compared to other macroscopic entities, like the wall of the vacuum recipient. No matter where the photon is absorbed, the absorption will result in an irreversible correlation of the ion with its environment, thus destroying the ion's coherences. However, the absorption in the wall does not yield macroscopically distinct states in the sense that an observer could access the information on the ion's state stored in the post-absorption state of the wall (as opposed to the case when the photon hits the detector.) Should the ion's state reduction (here to state $|1\rangle$) only happen if an apparatus yields distinct read-outs, then this would mean that the ion's dynamics depends on whether the photo detector is switched on or off during a sequence of $N$ measurements (which seems implausible.) Such a sequence of measurements (only the last one of $N$ measurement results is actually registered) has not been performed experimentally with a single ion. One would assume that the ion's dynamics is not changed during such a sequence compared to one where all intermediate results are "amplified" to distinct read-outs (as was actually done, and is necessary to demonstrate the quantum Zeno effect.) If this assumption is correct, then this together with the experimental results described below, implies that state reduction of the ion occurs independently of the information gain of any observer. In addition, it means that after one photon has been scattered by the ion, the ion is in state $|1\rangle$. Once it is in state $|1\rangle$, it will scatter many more photons during the detection interval (about $10^7 s^{-1}$), some of which will be registered by the photo detector. Therefore, if and only if the ion is in state $|1\rangle$, will a distinctive macroscopic read-out be obtained (resulting from photo detection) corresponding to this state. Similar reasoning shows that the absence of photo counts correlates with the ion being in state $|0\rangle$.

### 4.3.2 Fractionated $\pi$-pulse

To demonstrate the quantum Zeno paradox we investigate the impediment of an induced transition by means of measurements, similar to the proposal in (Cook 1988). First, an $^{171}Yb^+$ ion is illuminated for 50ms with uv light and thus prepared in state $|0\rangle$. The intensity, detuning and duration of microwave radiation applied to the ion is adjusted such that a $\pi$-pulse results, inducing a transition to state $|1\rangle$. This is achieved, if subsequent probing using uv light invariably leads to registration of fluorescence yielding an "on" result (Fig. 6a). The duration of the $\pi$-pulse was 2.9 ms.

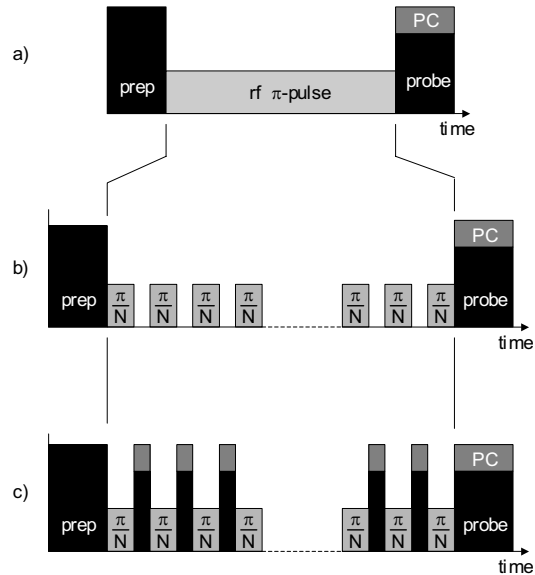

Figure 6: a: Excitation of the hyperfine transition in $^{171}Yb^+$ by applying a microwave $\pi$-pulse. b: Fractionated $\pi$-pulse without intermediate probing, and c: with intermediate probing using light at 369nm and simultaneous detection of the scattered photons (PC: photon counter, prep: initial preparation by a light pulse at 369nm.)

In order to separate the influence of the measurement pulses clearly from the driving field, the applied $\pi$-pulse is fractionated in $N$ pulses of equal area $\pi/N$ a time $\tau_{\text{probe}} = 3$ms apart (Fig. 6b). The frequency of microwave radiation is carefully set to resonance with the ionic transition by means of Ramsey-type experiments (compare section 2.) Thus, there is no dephasing between driving field and ion due to free precession during the intermissions, and the fractionated excitation will again result in a transition with nutation angle $N \times \pi/N$. Pulses of probe light are applied during the $N-1$ "gaps" of duration $\tau_{\text{probe}}$, and the photon counter is gated open synchronously in order to register or not register scattered photons indicating the ions excitation to state $|1\rangle$ or survival in the initially prepared state $|0\rangle$, respectively (Fig. 6c.) The experimental succession of initial preparation in $|0\rangle$, applying $N$ microwave pulses and $N$ probe pulse is repeated $2000/N$ times.

The experiment is carried out for $N = 1, 2, 3, 4$ and 10. We are interested in those sequences where all $N$ measurements give a negative ("off") result, indicating the survival of the ion in its initially prepared state. The number of these sequences normalized by the total number of sequences is plotted in Figure 7 versus the number of probe interventions (grey bars.) The data shown in Figure 7 have been corrected to account for the imperfect initial state preparation with an efficiency of 82%, as well as possible false detection of one of the $N$ results. The number of photo counts during a detection interval are Poisson distributed characterized by mean photon numbers of about 5 ("on" results) and 0.2 ("off" result), respectively. Since the two distributions overlap to some degree, wrong assignments may occur. To distinguish between "on" and "off" a fixed threshold is used. This threshold is chosen such that in less than 0.5% of the cases an "on" result is mistaken as "off". The error bars represent the variance of the binomial distribution of the number of recorded sequences of "on" and "off" results. In contrast to the proposal by Cook (Cook 1988) the result of each of $N$ measurements is registered. Therefore, it is possible to identify sequences of results that represent survival of the ion in the initially prepared state, $|0\rangle$ during the $N$ observations. The survival probability vanishes for $N = 1$ and increases to 77% for $N = 10$ showing that the evolution is impeded by frequent measurements.

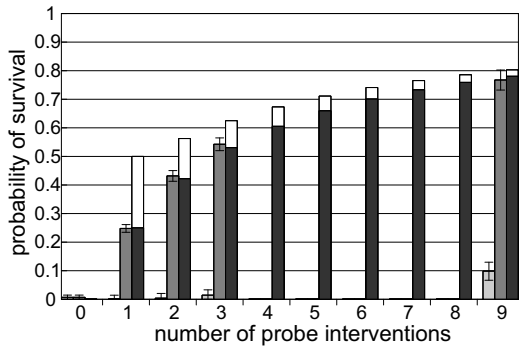

Figure 7: Probability of survival in the initially prepared state versus the number, $N-1$ of probe interventions. The gray bars indicate the corrected (see text) measurement results which agree well with the calculated values of the survival probability $P_{00}$ (black bars.) This demonstrates the quantum Zeno paradox. The light grey bars give the survival probability when no measurement pulses are applied. The measured values differ significantly from the values of the probability $P_{e0}$ (white bars,) that doesnot properly describe the quantum Zeno effect (see text.)

The occurrence of sequences of $N$ equal results ("off") follows $P_{00}(N) = \cos^{2N}(\pi/2N)$, according to equation (19). This is different from the "net" probability $P_{e0}(N) = 1/2(1 + \cos^N(\pi/N))$ where intermediate transitions between $|0\rangle$ and $|1\rangle$ are taken into account as discussed above. These two probabilities are significantly different for small values of $N$. The quantum Zeno paradox is evident in the correspondence of the experimental data in Figure 7 with $P_{00}$.

In principle it is possible to analyze the recorded data by ignoring the results of the first $N-1$ probe interventions in each sequence. More specifically, one could extract from the data the probability for the ion to end up in state $|0\rangle$ after the $N$-th measurement *irrespective* of its history. This probability would then be expected to agree with results from an ensemble experiment, provided no dephasing in the ensemble occurred. However, owing to the population accumulating in the Zeeman sublevels $|F = 1, m_F = \pm 1\rangle$ during an "on" detection interval, the upper state $|1\rangle$ may be decoupled by the probe light from the two-level system (Balzer, Hannemann, Reiß, Neuhauser, Toschek & Wunderlich 2002a): once an "on" result has been obtained, the ion may have made a transition to one of the Zeeman levels $|F = 1, m_F = \pm 1\rangle$ and is no longer affected by the subsequent microwave driving pulse. (This does not affect the determination of the survival probability in state $|0\rangle$.)

### 4.3.3 Statistics of sequences of equal results

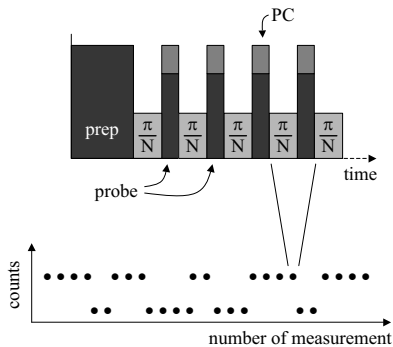

Figure 8: Top: Measurement scheme of alternating excitation and detection. Bottom: schematic of a trajectory of results.

In the experiment described in the previous section, the quantum Zeno effect is manifest in the survival probability of the initially prepared state $|0\rangle$ growing with the number of intermediate measurements during the driving $\pi$-pulse. In other words, frequent measurements hinder the transition to state $|1\rangle$, in accordance with Cook's suggestion to demonstrate the quantum Zeno effect. However, to demonstrate the quantum Zeno effect, it is not necessary to employ a fractionated $\pi$-pulse. The retardation of the evolution of an initially prepared state will show up in a sequence of alternated driving and probing, too. We have recorded series of 10000 pairs each consisting of a drive pulse and a probe pulse as shown in Figure 8(top) resulting in trajectories of alternating sequences of "on" and "off" results (Fig. 8, bottom). The normalized number of sequences of $q$ equal results, $U(q)$ corresponds, to good approximation, to the probability of survival in one of the eigenstates, $P_{00}$:

$$U(q)/U(1) = P_{00}(q-1) , \qquad (24)$$

where $U(1)$ denotes the probability to find the ion in this state at the very beginning of a sequence. $P_{00}$ is the conditional probability according to equation 19. The statistical distributions of the "off" sequences, $P_{00}(q-1) = U(q)/U(1)$ is shown in Figure 9 for the nutation angles $\theta = \pi, \pi/2, \pi/5$ and $\theta = 2\pi - 0.1$ (dashed lines). The duration of a microwave pulse that corresponds to $\theta = 2\pi$ was set here to 4.9 ms. The interval of probing was 2ms. The solid lines indicate the survival probabilities $P_{00}(q-1)$. A systematic deviation of 3% from the preset values of $\theta$ emerges from slightly varying the preset areas of the driving pulse while fitting $P_{00}$ to the data. For long sequences (large $q$), the data show strong deviations from the calculated survival probability. This is due to the finite length of the experimental trajectory.

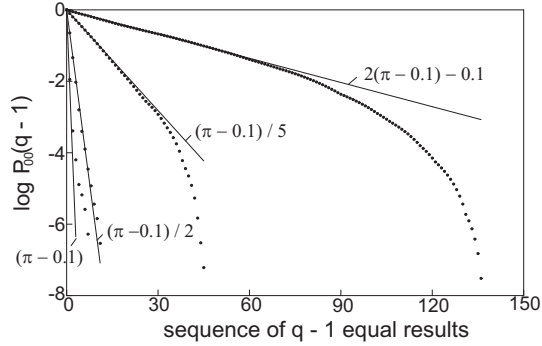

Figure 9: Statistical distribution of "off" sequences, $P_{00}(q-1) = U(q)/U(1)$ versus $q-1$ (for different values of the nutation angle $\theta$ indicated in the Figure.) The correspondence of the measurement results (dots) with the survival probabilities $P_{00}(q-1)$ (solid lines) verifies the quantum Zeno effect. Deviations between measured and calculated values for large $q$ are due to the finite length of the experimental trajectories of measurement results.

Both experiments on the hyperfine transition of the $^{171}$Yb+-ion show clearly the quantum Zeno effect, the retardation of the evolution of an individual quantum system as a consequence of measurements. In particular, they demonstrate the quantum Zeno *paradox*, since the measurement results are of the negative-result type, indicating a correlation between the observed individual quantum system and the measurement apparatus without local physical interaction.

# 5 Quantum state estimation using adapative measurements

## 5.1 Introduction

Determining an arbitrary state of a quantum system is a task of central importance in quantum physics, and in particular, in quantum information processing and communication where quantum mechanical 2-state systems (qubits) are elementary constituents. In order to gain complete knowledge about the state of a quantum system infinitely many measurements have to be performed on infinitely many identical copies of this quantum state. Naturally, the question arises how much information about a quantum state can be extracted using finite resources and what strategies are best suited for this purpose. A first indication of the appropriate operations to be carried out with two identically prepared qubits in order to gain maximal information about their state was given by (Peres & Wootters 1991). It was strongly suggested that optimal information gain is achieved when a suitable measurement on both particles together is performed. The measure that served to quantify the gain of information in these theoretical considerations was the Shannon information.

The Shannon information (or entropy) $-\sum_n p_n \log_2 p_n$ (Shannon 1948) is a measure for the uncertainty about the true value of some variable before a measurement of this variable takes place. The variable may take on $m$ different values with probability $p_n$, $n = 1..m$. Alternatively, the Shannon information can be viewed as giving a measure for the information that is gained by ascertaining the value of this variable. In Ref. (Brukner & Zeilinger 2001) it is argued that this measure is not adequate in the quantum domain, since the state of a quantum system is not well defined prior to observation. Only if the quantum system is in an eigenstate before and after a measurement is performed, does the measurement indeed reveal a preexisting property. In general, however, this is not the case in the quantum domain, and therefore, the Shannon information is deemed not suitable as a measure for the uncertainty associated with an observable before a measurement takes place. This statement may also be expressed in different words: In the quantum world not even the possible alternatives of measurement outcomes are fixed before a measurement is carried out. This fact is also at the core of the EPR programme where quantum mechanics predicts nonclassical correlations between two particles (Einstein et al. 1935). An alternative measure for the information content of a quantum system invariant under unitary transformations has been suggested in (Brukner & Zeilinger 1999).

The suggestion in (Peres & Wootters 1991) that the optimal measurement for determining a quantum state of two identically prepared particles needs to be carried out on both particles together, was proven in (Massar & Popescu 1995). In more technical terms this means that the operator characterizing the measurement does not factorize into components that act in the Hilbert spaces of individual particles only. In (Massar & Popescu 1995) it was also shown that the same is true when $N = 1, 2, 3, \ldots$ identically prepared qubits are available. The states to be estimated were drawn randomly from a uniform distribution over the Bloch sphere and the cost function that has been optimized was the fidelity $\cos^2(\theta/2)$ where $\theta$ is the angle between the actual and estimated directions. The optimal fidelity that can be reached is $(N + 1)/(N + 2)$. As a special case of optimal quantum state estimation of systems of arbitrary finite dimension the upper bound $(N + 1)/(N + 2)$ for the mean fidelity of an estimate of $N$ qubits was rederived in Ref. (Derka et al. 1998). In addition, it was shown that *finite* positive operator valued measurements (POVMs) are sufficient for optimal state estimation. This result implied that an experimental realization of such measurements is feasible, at least *in principle*. Subsequently, optimal POVMs were derived to determine the pure state of a qubit with the *minimal* number of projectors when up to $N = 5$ copies of the unknown state are available (Latorre et al. 1998). Still, the proposed optimal and minimal strategy requires the experimental implementation of rather intricate non-factorizing measurement operators. In addition, all $N$ qubits have to be available simultaneously for a measurement.

In (Bužek et al. 1999, Bužek et al. 2000) investigations are reported on how an arbitrary qubit state $|\psi\rangle$ can be turned into the state $|\psi^\perp\rangle$ orthogonal to the initial one. Such a quantum mechanical universal NOT (U-NOT) operation would correspond to the classical NOT gate that changes the value of a classical bit. It is shown that a U-NOT gate corresponds to an anti-unitary operation, and an ideal gate transforming an unknown quantum state into its orthogonal state does not exist. If a single qubit in a pure state is given and no *a priori* information on this state is available, then measuring the quantum state and using this information to prepare a state $|\psi^\perp\rangle$ gives the optimal result. If $N$ qubits in state $|\psi\rangle$ are available, then too, the optimal U-NOT operation can be attained by estimating the initial quantum state using these $N$ qubits and subsequently preparing the desired state. Thus, the optimal fidelity $(N + 1)/(N + 2)$ for a U-NOT is reached which coincides with the optimal fidelity for state estimation.

(Gill & Massar 2000) consider the problem of quantum state reconstruction when taking advantage of a large ensemble of identically prepared quantum states in a finite dimensional Hilbert space. For $N \to \infty$ any sensible measurement strategy yields a perfect estimate of a given quantum state, and since for large $N$ the estimate drawn from any strategy comes very close to the true value, the distinguishing feature between different strategies applied to large ensembles is the *rate* at which neighboring states can be distinguished. A quantitative measure for this rate is introduced and an upper bound for any type of estimation strategy is derived in (Gill & Massar 2000). For the case of a 2-dimensional Hilbert space (qubits) an explicit measurement strategy for pure states is given attaining this upper bound when using separate measurements on each particle. It turned out that for mixed states this upper bound is also valid as long as measurements are carried out in a factorizing basis. However, if collective measurements are allowed for, then this bound is not necessarily valid. Therefore, mixed states exhibit nonlocality without entanglement when large ensembles are available whereas pure states do not show this feature.

Nonlocality without entanglement has been described in Ref. (Bennett et al. 1999). There *unentangled* quantum states of a composite quantum system are described that can only be distinguished by a joint measurement on the whole system, but not by separate measurements on the individual constituents, not even when exchange of classical information between the observers measuring the individual objects is allowed for. A joint measurement on the quantum system reveals more information than any "classically" coordinated measurements of the individual parts.

In (Massar & Popescu 2000) it is shown that different definitions for the tar-get function function that is to be maximized in quantum state estimation may lead to different recipes for optimal measurements. This in turn will determine the properties of the quantum state that is revealed in a quantum measurement. It is shown that no matter what type of target function is chosen, the maximum amount of information that can be obtained from one qubit is one bit.

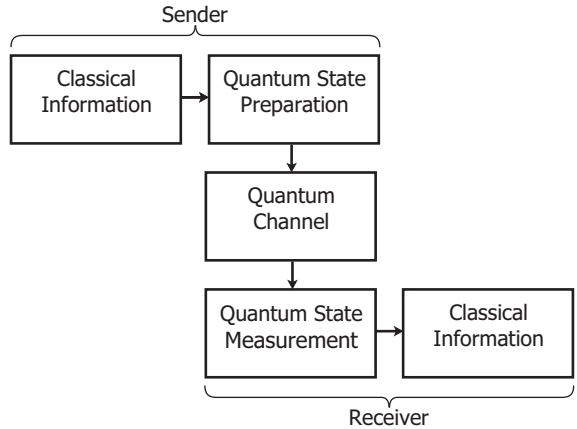

Figure 10: The steps necessary for transmission of information using quantum systems are schematically shown.

Estimating a quantum state can also be viewed as the decoding procedure at the receiver end of a quantum channel necessary to recover elements of an alphabet that have been encoded in quantum states by a sender (see, for instance, (Jones 1994).) A sketch of the steps necessary for the transmission of quantum information is displayed in Figure 10: The sender prepares a quantum state by setting the classical parameters of an appropriate device that prepares the desired quantum state. Then, the quantum system propagates in space or time until it reaches the receiver equipped with an apparatus capable of performing measurements in any basis, and it is her/his task to give the best possible estimate of this quantum state after $N$ identically prepared copies of the quantum system have been sent. In order to specify what "best possible" means, the Shannon information and von Neumann entropy for this situation have been computed, and an upper bound for the information obtainable from $N$ identically prepared quantum states as well as a lower bound on the entropy have been derived (Jones 1994).

The quantum information associated with a state of a qubit to be transmitted can be viewed as a unit vector indicating a direction in space. If no common coordinate system has been established, then the transmission of a direction in space between two distant parties requires a physical object. In the quantum domain, $N$ identically prepared spin-1/2 systems may serve for this purpose. It has been shown that the optimal state, that is, the one that yields the highest average fidelity $F = \langle \cos^2(\theta/2) \rangle$ of transmission is an entangled one for $N > 2$ (Bagan et al. 2000, Peres & Scudo 2001, Bagan, Baig, Brey & Munoz-Tapia 2001) ($\theta$ is the angle between the estimated and the actual direction to be transmitted.) The use of product states for communication of a spin direction has been investigated in (Bagan, Baig & Munoz-Tapia 2001).

Debugging of a quantum algorithm is another possible application for quantum state estimation (Fischer et al. 2000). Once a quantum algorithm has been implemented, it has to be tested, for instance by checking the state of a certain qubit in the course of the computation. In such a case the qubits are only available sequentially and efficient estimation is desirable, that is, a large overlap of the estimated state with the true one while keeping the number of repetitions of the algorithm as small as possible.

First experimental steps towards entanglement-enhanced determination ($N = 2$) of a quantum state have been undertaken (Meyer et al. 2001). The rotation angle around a specific axis of the total angular momentum of 2 spin systems has been estimated with an uncertainty below the standard quantum limit. The related problem of measuring in an optimal way the phase difference $\phi$ between the two arms of a Mach-Zehnder interferometer has been addressed in Ref. (Berry & Wiseman 2000). The optimal input state has been derived and an adaptive measurement scheme is proposed that relies on the detection of photon counts and yields a variance in $\phi$ close to the optimal result.

We have seen that optimal strategies to read out information encoded in the quantum state of a limited number of identical qubits require intricate measurements using a basis of entangled states. The first experimental demonstration of a self-learning measurement (employing a factorizing basis) of an arbitrary quantum state (Hannemann et al. 2002) and an experimental comparison with other strategies is reviewed in what follows.

## 5.2  Elements of the theory of self-learning measurements

It was recently shown that quantum state estimation of qubits with fidelity close to the optimum is possible when a self-learning algorithm is used (Fischer et al. 2000). When using this algorithm, $N$ members of an ensemble of identically prepared quantum systems in a pure state,

$$|\psi\rangle = |\psi\rangle_1 \otimes |\psi\rangle_2 \otimes \ldots \otimes |\psi\rangle_N \;, \tag{25}$$

can be measured individually, that is, they do not have to be available simultaneously. In other words, the measurement operator $\hat{M}$ employed to estimate the state can be written as a tensor product and we have

$$\hat{M}|\psi\rangle = \hat{m}_1|\psi\rangle_1 \otimes \hat{m}_2|\psi\rangle_2 \otimes \ldots \otimes \hat{m}_N|\psi\rangle_N \tag{26}$$

The operators $\hat{m}_n$ project onto the orthonormal basis states

$$|\theta_m^{(n)}, \phi_m^{(n)}\rangle = \cos\frac{\theta_m^{(n)}}{2}|0\rangle + \sin\frac{\theta_m^{(n)}}{2}e^{i\phi_m^{(n)}}|1\rangle \text{ and } |\pi - \theta_m^{(n)}, \pi + \phi_m^{(n)}\rangle \tag{27}$$

An experimental realization of a self-learning measurement on an individual quantum system in order to estimate its state is reported in (Hannemann et al. 2002). The projector $\hat{m}_n$ of measurement $n$ is varied in real time during a sequence of $N$ measurements conditioned on the results of previous measurements $\hat{m}_l, l < n$ in this sequence (for the first measurement, $n = 1$, obviously no prior knowledge of the state is available and the first measurement basis can be chosen arbitrarily.) The cost function that is optimized when proceeding from measurement $n$ to $n+1$ is the fidelity of the estimated state after measurement $n+1$. This will be detailed in the following paragraphs.

Prior to the first measurement no information on the qubit state is available and the corresponding density matrix $\varrho_0$ reflecting this ignorance is

$$\varrho_0 = \int_0^\pi d\theta \sin\theta \int_0^{2\pi} d\phi\, w_0(\theta,\phi)\, |\theta,\phi\rangle\langle\theta,\phi|\,, \tag{28}$$

where $w_0(\theta,\phi) = \frac{1}{4\pi}$ is the probability density on the Bloch sphere. After the qubit has been measured in the direction $(\theta_m, \phi_m)$ the new distribution $w_n(\theta,\phi)$ ($n=1,2,3,\ldots,N$) is obtained from Bayes rule ((Bayes 1763), reprinted in (Bayes 1958):

$$w_n(\theta,\phi|\theta_m,\phi_m) = \frac{w_{n-1}(\theta,\phi)\, |\langle\theta_m,\phi_m|\theta,\phi\rangle|^2}{p_n(\theta_m,\phi_m)}\,, \tag{29}$$

where $w_{n-1}(\theta,\phi)$ gives the *a priori* probability density to find the qubit along the direction indicated by $\theta$ and $\phi$. The conditional probability to measure the qubit along the direction $|\theta_m,\phi_m\rangle$ if it is in state $|\theta,\phi\rangle$ is given by $|\langle\theta_m,\phi_m|\theta,\phi\rangle|^2$. Correct normalization is ensured by the denominator

$$p_n(\theta_m,\phi_m) = \int_0^\pi d\theta \sin\theta \int_0^{2\pi} d\phi\, w_{n-1}(\theta,\phi)\, |\langle\theta_m,\phi_m|\theta,\phi\rangle|^2 \tag{30}$$

that gives the probability to measure the qubit along the direction $|\theta_m,\phi_m\rangle$ irrespective of its actual state, that is, integrated over all possible *a priori* directions.

The adaptive algorithm needs to find optimal measurement axes $(\theta_m,\phi_m)_n$ after each step. The optimization is based on the knowledge gained from the preceding measurements as represented by $w_{n-1}(\theta,\phi)$.

The cost function used to find the optimal measurement is the fidelity

$$F_{n-1}(\theta,\phi) = \langle\theta,\phi|\varrho_{n-1}|\theta,\phi\rangle\,. \tag{31}$$

After $n-1$ measurements the knowledge of the state is represented by $w_{n-1}(\theta,\phi)$ and the fidelity of any state $|\theta,\phi\rangle$ is

$$F_{n-1}(\theta,\phi) = \int_0^\pi d\theta' \sin\theta' \int_0^{2\pi} d\phi'\, w_{n-1}(\theta',\phi')\, |\langle\theta,\phi|\theta',\phi'\rangle|^2 \tag{32}$$

The estimated state after $n-1$ measurements $|\theta_{\text{est}},\phi_{\text{est}}\rangle_{n-1}$ has to maximize this fidelity:

$$F_{n-1}(\theta_{\text{est}},\phi_{\text{est}}) = F_{n-1}^{\text{opt}} \equiv \max F_{n-1}(\theta,\phi) \tag{33}$$

In order to find the optimal direction for the next measurement $n$, the *expected* mean fidelity after the next measurement is maximized as a function of the measurement axis. Suppose the system will be found in the direction $(\theta_m,\phi_m)$. Then the fidelity would be

$$F_n(\theta,\phi|\theta_m,\phi_m) = \int_0^\pi d\theta' \sin\theta' \int_0^{2\pi} d\phi'\, w_n(\theta',\phi'|\theta_m,\phi_m)\, |\langle\theta,\phi|\theta',\phi'\rangle|^2 \tag{34}$$

where the *expected* distribution *after* this measurement, $w_n(\theta', \phi'|\theta_m, \phi_m)$ is obtained from Bayes rule (Eq. 29). The optimal fidelity $F_n^{\text{opt}}(\theta_m, \phi_m)$ is then found by maximizing this function with respect to $(\theta, \phi)$.

A measurement along a certain axis will reveal the system to be in one of two possible states: Either is is found along this axis $(\theta_m, \phi_m)$, or in the opposite direction $(\bar{\theta}_m, \bar{\phi}_m) \equiv (\pi - \theta_m, \pi + \phi_m)$. So far we have only taken into account the first of these two possible outcomes of the measurement. The optimized fidelity for the second result is calculated analogously and occurs with probability $p_n(\bar{\theta}_m, \bar{\phi}_m)$ (Eq. 30). Thus the expected mean fidelity after the next measurement is given by the optimized fidelities for each outcome, weighted by the estimated probability for that outcome:

$$\bar{F}_n(\theta_m, \phi_m) = p_n(\theta_m, \phi_m) F_n^{\text{opt}}(\theta_m, \phi_m) + p_n(\bar{\theta}_m, \bar{\phi}_m) F_n^{\text{opt}}(\bar{\theta}_m, \bar{\phi}_m) \qquad (35)$$

The optimal measurement direction $(\theta_m^{\text{opt}}, \phi_m^{\text{opt}})$ has to maximize this function.

The direction of the first ($n = 1$) measurement is of course arbitrary, since there is no *a priori* information on the state. The expected mean fidelity in this case is $\bar{F}_1 = 2/3$, independently of $(\theta_m, \phi_m)_1$. For measurements two and three the following analytical expressions have been derived: $\bar{F}_2 = (1/2 + \cos(\alpha/2 - \pi/4)/\sqrt{18})$, where the expected mean fidelity depends on the relative angle $\alpha$ between the second and the first measurement direction. Thus the optimal second measurement direction has to be orthogonal to the first one, yielding $\bar{F}_2^{\text{opt}} = (1/2 + 1/\sqrt{18})$. The optimal third measurement axis is orthogonal to both previous directions and yields $\bar{F}_3^{\text{opt}} = (1/2 + 1/\sqrt{12})$.

Interestingly, if the Shannon information is used as a cost function to find the optimal measurement directions, then the fidelity obtained from numerical simulations of the estimated state after $N$ measurements is not as high as is the case, if the fidelity is employed as outlined above (Fischer et al. 2000). This may hint at the inadequacy of the use of the Shannon measure in the quantum domain as pointed out by Brukner and Zeilinger (Brukner & Zeilinger 2001). However, the $\log_2$ function occurring in the Shannon measure for information poses some difficulties when numerically optimizing the cost function, and the less precise final estimate of the quantum state in our numerical studies could be caused by accumulating round-off errors.

### 5.3 Experiment

A typical sequence of measurements where the adaptive algorithm outlined above has been realized is depicted in Figure 11. The probability density $w_n(\theta, \phi)$ is shown on the surface of the Bloch sphere and the measurement direction $n$ is indicated by a grey "pin. The head of the pin marks the outcome of the measurement along this direction. Here, the state to be estimated is $|\theta_{\text{prep}}, \phi_{\text{prep}}\rangle = |3\pi/4, \pi/4\rangle$. The white solid circle on the Bloch sphere represents the parameters $\theta_p$ and $\phi_p$ of the state to be estimated, and does not indicate a quantum mechanical uncertainty. These parameters are part of a recipe to prepare the desired quantum state using a classical apparatus. When such a quantum state is subjected to a measurement, for instance, along the $z$-direction, then after this measurement, of course, there will be no more information available about the components of the initial state in the $x$- and $y$-directions in accordance with the uncertainty relation derived from the commutators of the spin-1/2 operators. Only in the limit $N \to \infty$ for $N$ suitably chosen measurements of $N$ states prepared according to the same recipe, the parameters $\theta$ and $\phi$ could be recovered. The uncertainty associated with the preparation of a specific quantum state, $|\theta, \phi\rangle$ is not a quantum mechanical one, it is determined by technical issues: If the electromagnetic field used for preparation of an ionic quantum state contains a large number of photons, for example, a coherent intense field emitted by a mw source with mean photon number $\langle M \rangle$ satisfying $\langle \triangle M \rangle / \langle M \rangle \ll 1$ (Haroche 1971), then the "graininess" of the field can be safely neglected, and the amplitude stability of the applied mw field determined by technical specifications of the mw source would limit the precision of state preparation. The time resolution (25ns) of the digital signal processing system controlling the mw source is another technical limitation for the accuracy and precision of state preparation. (In the actual experiment, the initial preparation of state $|0\rangle$ is the main source of imprecision when an arbitrary quantum state is generated.)

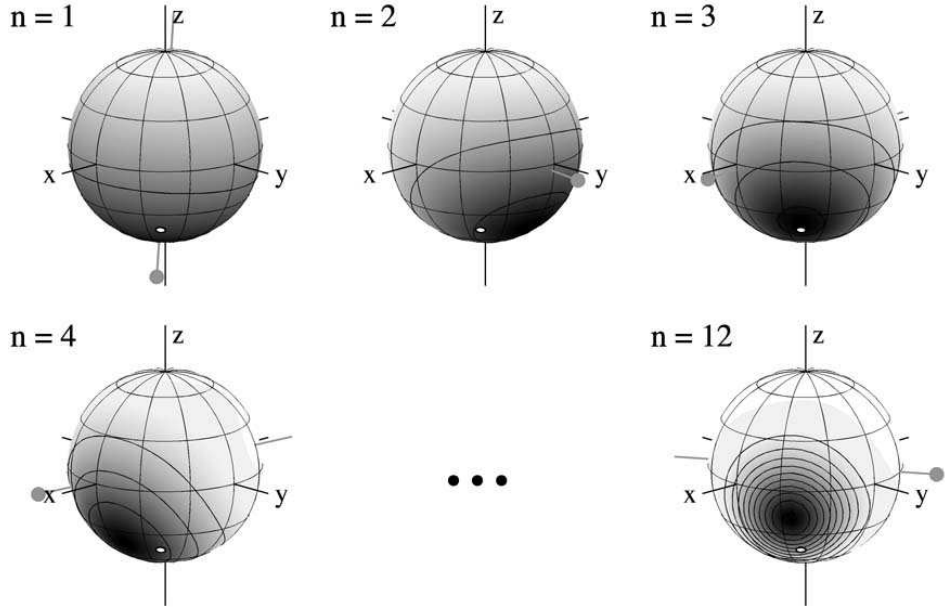

Figure 11: A sequence of $N = 12$ adaptive measurements carried out on identically prepared qubits in order to estimate their state ($|3\pi/4, \pi/4\rangle$, marked by a white solid circle.) The probability density $w_n(\theta, \phi)$ is gray scale coded on the surface of the Bloch sphere (the gray scale code is different for each measurement.) In addition, contour lines indicate where $w_n$ takes on the values $0.1, 0.2, \ldots$. A gray straight line through each Bloch sphere shows the measurement direction and the filled gray circle indicates the measurement outcome. The fidelity of state estimation of this particular run is 94.9%.

Imprecision in the initial preparation of $|0\rangle$ and in the subsequent preparation of a desired quantum state, relaxation and dephasing of the quantum state before it is being measured, and the effect of an imperfect measurement can be concisely summarized as the action of a depolarizing quantum channel together

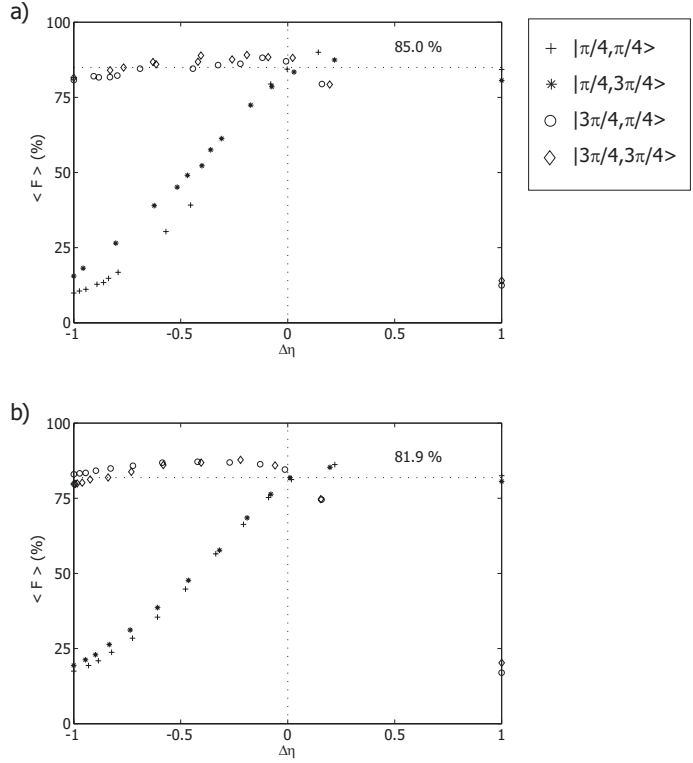

Figure 12: The experimentally determined average fidelity of state estimation as a function of the difference in detection efficiency, $\Delta\eta$ for states $|0\rangle$ and $|1\rangle$, respectively. The fidelity is plotted for different states to be estimated. For $\Delta\eta = 0$ the fidelity should be independent of the initial state which is indeed observed in the experiment. a) Self-learning estimation; b) Random choice of basis.

with a systematic bias.

$$\varrho \to (1 - 2\lambda)\varrho + \lambda I + \Delta\eta\, \sigma_z \;, \tag{36}$$

where $\Delta\eta \equiv (\eta_1 - \eta_0)/2$ is the difference in detection efficiencies for state $|1\rangle$ and $|0\rangle$, respectively; $0 \leq \lambda \leq 1/2$, and here we have $\lambda \leq 1 - \bar{\eta} \equiv 1 - (\eta_1 + \eta_0)/2$ with $1/2 \leq \eta_{0,1} \leq 1$. This description is also applicable to other types of experiments where imperfections may be due to other physical reasons. The third term on the rhs in eq. 36 arises whenever the efficiencies of detection for states $|0\rangle$ and $|1\rangle$ ($\eta_0$ and $\eta_1$, respectively) differ from each other, and has a specific influence on different estimation strategies. For any strategy, $\Delta\eta \neq 0$ means that the fidelity of state estimation depends on the state to be measured as can be seen in Fig. 12. Fig. 12 also displays data obtained from experimental runs where the $N$ measurement directions are chosen randomly.

Experiments are necessarily imperfect, that is, they never perfectly reflect results obtained from theoretical considerations. In the case of quantum state estimation this means that an estimate with fidelity equal to the theoretical value cannot be obtained. Here, the performance of the experimental apparatus has been characterized quantitatively and completely (that is, the features that are relevant for the experiment.) Taking into account the known experimental imperfections, the theoretical value for the fidelity of state estimation is numerically calculated for an ensemble of 10000 states drawn randomly from a uniform distribution on the Bloch sphere. This theoretical mean fidelity is then compared to the experimental result of the self-learning algorithm and the random strategy (Fig. 13.)

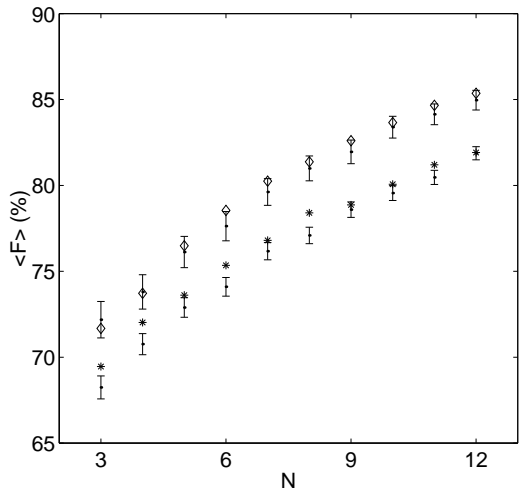

Figure 13: The average fidelity of state estimation as a function of the number $N$ of available qubits. Diamonds and stars indicate values obtained from numerical simulations for the self-learning algorithm and the random choice of measurement basis, respectively (taking into account the experimental preparation and detection efficiencies.) Solid squares show the experimentally determined values for these two measurement strategies.

Decoherence inevitably occurs in any experiment and it has been shown that under this commonplace condition the self-learning strategy still yields the best results. Even more, the self-learning and the random strategy show a larger difference in mean fidelity (85.0% compared to 81.9%, the difference exceeding 5 standard errors) in the 'real' experimental world than the difference between the ideal theoretical values (92.5% and 91%, respectively, for $N = 12$.)

The estimation procedure discussed here allows for separate (local) measurements on each qubit. Following each measurement on a particular qubit, classical information is used to determine the best measurement to be performed on the next qubit. In reference (Bagan et al. 2002) the optimal LOCC scheme (performing local operations with exchange of classical information) is introduced for arbitrary states on the Bloch sphere (3D case). Interestingly, if the state to be estimated lies in the $xy-$plane (2D case), then local operations alone suffice to obtain the optimal state estimate and classical communication is not necessary. This optimal LO(CC) scheme exhibits the same asymptotic behavior with the number $N$ of qubits as the optimal scheme taking advantage of collective measurements, and yields (according to theory) a slightly better average fidelity than the adaptive scheme presented here.

# 6 Quantum Information

The investigations of fundamental questions of quantum mechanics, in addition to their intrinsic interest, may also prove useful to construct an information processor functioning according to the principles of quantum mechanics. If such a quantum computer were available, it could be used for a variety of tasks a classical computer, for all practical purposes, could not handle. A famous example for such a task is the factoring of large numbers. The difficulty of this task ensures the security of communication encrypted according to the RSA procedure (developed by R. Rivest, A. Shamir, and L. Adleman (Rivest et al. 1978)), if the large number the encryption is based on, is changed after a time interval short compared to the computational time needed to carry out the factorization of this number. In (Shor 1994, Shor 1997) an algorithm is described, based on the laws of quantum mechanics, that could be used to find the prime factors of a given number of the order $10^N$ in time $T$ proportional to $(\ln N)^3$, as compared to the best known classical algorithm where $T \propto \exp(s(\ln N)^{1/3}(\ln \ln N)^{2/3})$ with $O(s) = 2$ .

For $N = 130$ (a typical order of magnitude of today's RSA-type encrypting schemes) a classical computer using the best known algorithm would need about $10^{17}$ instructions, whereas a quantum computer running Shor's algorithm (Shor 1994) could solve the same task using $10^{10}$ instructions, thus reducing the time needed for this computation by a factor $10^7$, if the same speed for an elementary operation is assumed for both types of computers. The efficiency of the Shor algorithm relies on the efficient implementation of Fourier transformations with quantum logic. Its advantage over the classical algorithm increases with increasing complexity of the problem to be solved (here with $N$.) It was shown, that a single qubit in a pure state together with $\log_2 N$ qubits in arbitrary mixed states suffice to implement Shor's factoring algorithm (Parker & Plenio 2000).

The calculation of properties or dynamics of quantum systems is a promising line of action for a quantum computer, even one with only few qubits and operating with limited precision (Feynman 1982, Mølmer & Sørensen 2000). In Ref. (Abrams & Lloyd 1999) it is shown how a quantum computer consisting of about 100 qubits can be used to calculate eigenvalues and eigenvectors of Hamilton operators. Computing, for instance, energy levels and correlation energies of a Boron atom with 5 electrons is a rather intricate problem: if 20 angular wave functions and 40 radial wave functions are used, then this amounts to a total of about $10^{15}$ many body basis states to be considered in such a calculation. Sophisticated classical techniques have been developed to circumvent problems arising from the exponentially growing space of basis states. Still, a quantum computer of very limited size may be able to perform more accurate calculations (Abrams & Lloyd 1999). Ref. (Somaroo et al. 1999) describes how proton nuclear spins have been used to simulate the dynamics of a truncated quantum harmonic oscillator employing nuclear magnetic resonance techniques. Nonlinear dynamical problems that are hard or impossible (for all practical purposes) to solve on a classical computer due to accumulating round-off errors may also be simulated efficiently on a quantum computer (Georgeot & Shepelyansky 2001a, Georgeot & Shepelyansky 2001b, Georgeot & Shepelyansky

2002).

What is the origin of the computational power of a quantum computer? The elementary switching unit (bit) of usual classical computers is a transistor that may assume two distinct macroscopic states that can be identified with the computational binary states 0 and 1. In a quantum computer transistors are replaced by two-state quantum systems (qubits) that may exist in arbitrary superposition states $\alpha|0\rangle + \beta|1\rangle$ with the complex numbers $\alpha, \beta$ satisfying $|\alpha|^2 + |\beta|^2 = 1$. The possibility to exploit the quantum mechanical superposition principle and the linearity of operations in Hilbert space for massive parallel computing is one ingredient for a quantum computer. The art of designing quantum algorithms makes use of another feature of quantum mechanics: the ability to display interference. Roughly speaking, a quantum algorithm has to be designed such that different computational paths interfere in such a way that at the end of the algorithm the correct result survives with probability near unity(Cleve et al. 1998). Recent introductions to quantum computing can be found, for instance, in (Nielsen & Chuang 2000, Gruska 1999).

To date, nuclear magnetic resonance applied to macroscopic ensembles of molecules (Gershenfeld & Chuang 1997) and electrodynamically trapped ions (Cirac & Zoller 1995) are the two physical systems that have been most successfully used to demonstrate quantum logic operations, and even complete quantum algorithms (Jones & Mosca 1998, Chuang et al. 1998, Vandersypen et al. 2001). Also, their specific advantages and shortcomings have been thoroughly investigated, experimentally and theoretically. Introductions to quantum computing with an emphasis on ion traps or nuclear magnetic resonance are given, for instance, in (Steane 1997, Wineland et al. 1998, Sasura & Buzek 2002) and (Jones 2001), respectively.

Sections 5 and 6.1 describe experiments with trapped $^{171}$Yb$^+$ ions addressing basic questions of quantum mechanics that, at the same time, are relevant for QIP: the self-learning measurement of arbitrary qubit states and the realization and characterization of various quantum channels. These experiments also demonstrate the ability to perform arbitrary single-qubit gates with individual $^{171}$Yb$^+$ ions with high precision – a prerequisite for QIP. The coherence time of the hyperfine qubit in $^{171}$Yb$^+$ is, for all practical purposes, limited by the coherence time of microwave (mw) radiation used to drive the qubit transition.

In addition to single-qubit operations, a second basic ingredient is required for QIP with trapped ions: conditional quantum dynamics with, at least, two qubits. Any quantum algorithm can then be synthesized using these elementary building blocks (DiVincenzo 1995, Barenco et al. 1995). Communication between qubits, necessary for conditional quantum dynamics, is achieved via the vibrational motion of the whole ion string(Cirac & Zoller 1995, Sorensen & Molmer 2000, Jonathan et al. 2000). Thus, external (motional) and internal degrees of freedom need to be coupled. Driving a hyperfine transition with mw radiation (as in the experiments described in this article) does not allow for such a coupling, since the Lamb-Dicke parameter is essentially zero for long-wavelength radiation. Also, the inter-ion spacing in usual traps is much smaller than the wavelength of mw radiation and, therefore, individual addressing of ions is not possible. Section 6.2.2 describes how an additional magnetic field gradient applied to an electrodynamic trap individually shifts ionic qubit resonances thus making them distinguishable in frequency space. At the same time, coupling of internal and motional states is possible even for mw radiation. With the introduction of this additional static field, all optical schemes devised for QIP in ion traps can be applied in the mw regime, too.

Instead of applying usual methods for coherent manipulation of trapped ions, a string of ions in such a modified trap can be treated like a molecule in NMR experiments taking advantage of spin-spin coupling. A collection of trapped ions forms a $N$-qubit "molecule" with adjustable spin-spin coupling constants (second part of section 6.2.2.)

## 6.1 Realization of quantum channels

Quantum logic operations, and other experiments where coherent superpositions of quantum states have to remain intact for a certain time, are carried out ideally under perfect, noiseless conditions. However, the inevitable coupling of qubits to their environment and the imperfection inherent to any physical operation with qubits invariably degrade the performance of quantum logic operations.

A quantum channel describes the general dynamics of a qubit under propagation in space and/or time. This evolution of qubits can be associated with a physical device used to transmit quantum information (like an optical fiber). When employing quantum states to transmit information, the sequence of necessary steps can be visualized as follows (compare Figure 10): some physical apparatus is used to prepare a quantum state using a set of classical variables. The quantum state propagates, signified by the quantum channel until it is measured by a receiver, again using a suitable apparatus to extract the values of classical variables. The optimal reconstruction of quantum states has been the topic of experiments described in the previous section. Now we consider explicitly the influence of the environment on a quantum state once it has been prepared, that is, we investigate the influence of the quantum channel on the transmission of quantum information (Hannemann et al. n.d.).

The state of a qubit is completely determined by the expectation values $\langle\sigma_x\rangle, \langle\sigma_y\rangle$, and $\langle\sigma_z\rangle$, and the density matrix describing its state can be written as
$$\rho = \frac{1}{2}(I + \vec{s} \cdot \vec{\sigma}) \tag{37}$$
where $\vec{s} \cdot \vec{\sigma} = \langle\sigma_x\rangle\sigma_x + \langle\sigma_y\rangle\sigma_y + \langle\sigma_z\rangle\sigma_z$, and $\sigma_{x,y,z}$ are the Pauli matrices. Ideally, while being transmitted through the quantum channel, the qubit's state described by the Bloch vector $\vec{s}$ is not changed. However, in general, the propagation of the qubit through a quantum channel will alter the qubit's state and $\vec{s'}$ will be obtained at the quantum channel's exit. This change of the qubits state can be of reversible or irreversible nature. The quantum channel may also stand for a quantum memory storing a qubit state which may undergo some change until it is 'activated again, that is, transferred to another quantum system or being subjected to a measurement. It can also represent the dynamics of a qubit during a quantum computation. The most economical error correcting and avoiding codes used to correct or stabilize quantum information depend on the type of quantum channel the qubits are exposed to.

Two examples for detrimental effects acting on qubit states (the consequence of "noise") are given in what follows. A phase damping channel leads to decoherence of a qubit state, affecting the off-diagonal elements $\rho_{10} = \rho_{01}^* = \langle\sigma_x\rangle - i\langle\sigma_y\rangle$ of the density matrix that are diminished or disappear completely while the diagonal elements remain unchanged. It transforms a Bloch vector according to

$$\vec{s}\,' = \begin{pmatrix} 1-2\lambda & 0 & 0 \\ 0 & 1-2\lambda & 0 \\ 0 & 0 & 1 \end{pmatrix} \vec{s}\,, \tag{38}$$

with $0 \leq \lambda \leq 1/2$. The $\sigma_x$ and $\sigma_y$ components of the Bloch vector shrink by a factor $1-2\lambda$.

A quantum channel that fully depolarizes the quantum state of a qubit transforms any state $\rho$ into a completely mixed state $\rho' = 1/2\, I$. A partially depolarizing channel can be characterized by a parameter $0 \leq \lambda \leq 1/2$ that is interpreted as the probability for changing the qubit's state into its orthogonal state: if the input state is pure, then we choose the basis such that the qubit's initial density matrix reads

$$\rho = \frac{1}{2}(I + \sigma_z)\,. \tag{39}$$

After the quantum channel

$$\rho' = (1-\lambda)\frac{1}{2}(I + \sigma_z) + \lambda \frac{1}{2}(I - \sigma_z)\,. \tag{40}$$

The action of the depolarizing channel is independent of the initial polarization of the qubit, hence can be described by

$$\vec{s} \to \vec{s}\,' = (1-2\lambda)\vec{s}\,. \tag{41}$$

In (Fujiwara & Algoet 1999) it is shown that any quantum channel for qubits can be cast in the form

$$\vec{s}\,' = \hat{M}\vec{s} + \vec{v}\,. \tag{42}$$

where $\hat{M} \in \mathbb{R}^{3\times 3}$ and $\vec{v} \in \mathbb{R}^3$. Equation 42 yields $\vec{s}\,'$, the Bloch vector of the qubit after it has traversed the quantum channel characterized by $\hat{M}$ and $\vec{v}$.

Various quantum channels have been realized experimentally with $^{171}$Yb$^+$ ions (Hannemann et al. n.d.) and the matrix and vector elements

$$\begin{aligned} M_{ij} &= 2P_{ij} - P_{iz} - P_{i(-z)} \\ v_i &= P_{iz} + P_{i(-z)} - 1 \end{aligned} \tag{43}$$

are determined by measuring the probabilities (or rather relative frequencies) $P_{ij} = \langle i | \rho' | j \rangle$, where $\rho'$ is the density matrix describing the qubit state after the quantum channel, and $i, j \in \{x, y, z\}$ (Hannemann et al. n.d.).

Exploiting coherent and incoherent operations on the hyperfine qubits of Yb$^+$ we realized and completely characterized a polarization rotating quantum channel, a phase damping quantum channel acting in the $xy-$plane, and a phase damping quantum channel acting in an arbitrary plane. A Pauli channel and combinations of the aforementioned channels can also be realized. Incoherent disturbances to a quantum channel are realized by applying to the qubit a noisy magnetic field with well-defined spectral properties in conjunction with coherent microwave operations. Another possibility to produce a desired quantum channel is realized by applying to the qubit small amounts of light close to 369nm, thus inducing well-defined quantities of longitudinal and/or transversal relaxation during coherent microwave operations (Balzer, Hannemann, Reiß, Neuhauser, Toschek & Wunderlich 2002b). This light-induced decoherence is readily applicable to individually addressed quantum systems, it may be switched on and off immediately, and it is reproducible. Although, in the present experiment, the coherent drive was microwave radiation resonant with a ground-state hyperfine transition in Yb$^+$ , the same principle seems to apply to a system where a dipole-forbidden optical transition is driven by laser light (for example in Ca$^+$ or Ba$^+$ .)

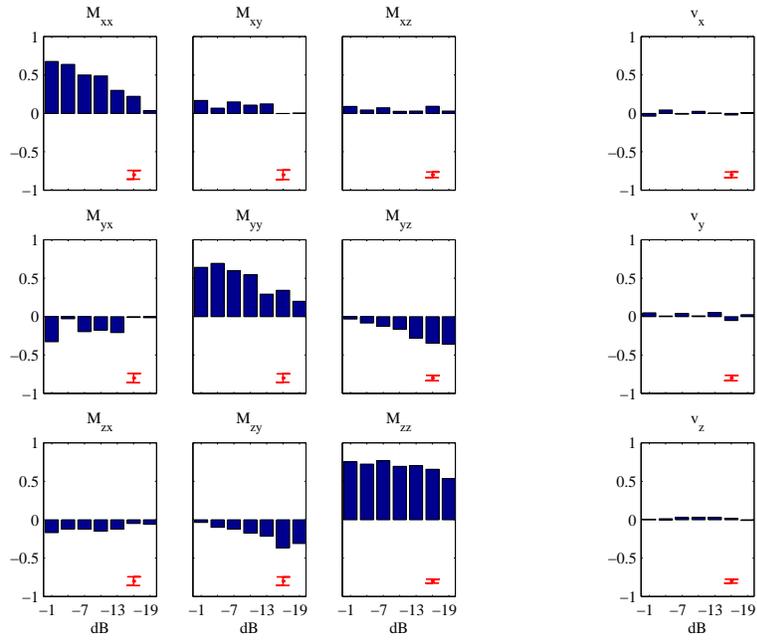

Figure 14: Experimentally realized quantum channel with designed phase damping in the plane normal to the unit vector $\vec{n} = (1, \pi/6, 0)^T$ (in polar coordinates). In addition, a small amount of amplitude damping is present. The relative amplitude of the noise magnetic field was varied between -19dB and -1dB.

An example of an experimentally realized quantum channel is displayed in Figure 14. Here, a channel affected by controlled amounts of phase damping in the plane normal to the unit vector $\vec{n} = (1, \pi/6, 0)^T$ (in polar coordinates) has been implemented. The applied noisy magnetic field causes in addition a small amount of amplitude damping (detailed in (Hannemann et al. n.d.).) Each matrix element is plotted as a function of the amplitude of the additional noisy magnetic field.

Since the elements of the matrix describing the quantum channel can be varied over a wide range, this experimental system can be used to simulate specific quantum channels characteristic for other physical implementations of QIP, too. Error correction is essential for QIP, since decoherence is ubiquitous. We have implemented different quantum channels characterized by reversible and irreversible dynamics that can be used, for instance, to experimentally test the capabilities of different types of quantum error correcting codes under varying conditions.

If spatially separated quantum information processors, for example, ion traps each containing a limited number of qubits are connected to allow for the exchange of quantum information, then it will also be useful to be able to first characterize the quantum channel and then apply the appropriate strategy to avoid or correct these specific errors.

## 6.2 New concepts for spin resonance with trapped ions

The discussion of a new approach to ion trap quantum computing in this section is restricted to the use of electrodynamic Paul traps, even though this concept should also be applicable when other trapping techniques are employed, for example, Penning traps (Powell et al. 2002).

### 6.2.1 Linear ion trap

In a linear Paul trap (Paul et al. Westdeutscher Verlag, Cologne,1958), a time-dependent two-dimensional quadrupole field strongly confines the ions in the radial direction yielding an average effective harmonic potential (Ghosh 1995). An additional static electric field is applied to harmonically confine the ions also in the axial direction (Prestage et al. 1989, Raizen et al. 1992). If the confinement of $N$ ions is much stronger in the radial than in the axial direction, the ions will form a linear chain (Schiffer 1993, Dubin 1993) with inter-ion spacing

$$\delta z \approx \zeta \, 2 N^{-0.56} \tag{44}$$

where

$$\zeta \equiv (e^2/4\pi\epsilon_0 m \nu_1^2)^{1/3} \, , \tag{45}$$

$m$ is the mass of one singly charged ion, $e$ the elementary charge, and $\nu_1$ is the angular vibrational frequency of the center-of-mass (COM) mode of the ion string (Steane 1997, James 1998). The distance between neighboring ions, $\delta z$ is determined by the mutual Coulomb repulsion of the ions and the trapping potential. Typically, $\delta z$ is of the order of a few $\mu$m; for example, $\delta z \approx 7\mu$m for $N = 10$ $^{171}$Yb$^+$ ions with $\nu_1 = 100 \times 2\pi$kHz.

Two appropriately chosen internal states of each ion confined in a linear electrodynamic trap represent a quantum mechanical 2-state system that may serve as one qubit. In order to prepare these quantum mechanical 2-state systems individually (single qubit operations), electromagnetic radiation is aimed at on ion at a time, that is, it must be focused to a spot size much smaller than $\delta z$. Therefore, optical radiation is usually required for individual addressing of qubits in ion traps (Nägerl et al. 1999).

In order to implement conditional quantum dynamics with ionic qubits, it is necessary (in addition to single qubit operations) to couple external and internal degrees of freedom. The interaction Hamiltonian governing the dynamics of a particular ion $j$ at position $z_j$ subjected to an electromagnetic field with angular frequency $\omega$ and initial phase $\phi'$ reads

$$\begin{aligned} H_I &= \frac{\hbar}{2}\Omega(\sigma_j^+ + \sigma_j^-)\left[\exp[i(kz_j - \omega t + \phi')] + \exp[-i(kz_j - \omega t + \phi')]\right] \\ &= \frac{\hbar}{2}\Omega(\sigma_j^+ + \sigma_j^-)\left[\exp\left[\sum_n^N iS_{nj}\eta_n(a_n^\dagger + a_n) - i\omega t + i\phi\right] + \text{h.c.}\right] . \end{aligned} \quad (46)$$

where $\Omega = \vec{d}\cdot\vec{F}/\hbar$ is the Rabi frequency with $\vec{d}\cdot\vec{F}$ signifying either magnetic or electric coupling between the atomic dipole and the respective field component. $\sigma_{+,-} = 1/2\,(\sigma_x \pm \sigma_y)$ represent the atomic raising and lowering operators, respectively, ($a_n^\dagger$ and $a_n$ are the creation and annihilation operators of vibrational mode $n$, and $S_{nj}$ are the coefficients of the the unitary transformation matrix that diagonalizes the dynamical matrix describing the axial degrees of freedom of a linear string of $N$ ions (Wunderlich 2001). The Lamb-Dicke parameters $\eta_n$ determining the coupling strength between internal and motional dynamics are given by

$$\eta_n \equiv \sqrt{\frac{(\hbar k)^2}{2m}/\hbar\nu_n} = \frac{\hbar k}{2\Delta p_n} = \frac{\Delta z_n\, 2\pi}{\lambda} . \quad (47)$$

The square of $\eta_n$ gives the ratio between the change in kinetic energy of the atom due to the absorption or emission of a photon and the quantized energy spacing of the harmonic oscillator mode characterized by angular frequency $\nu_n$. The mean square deviation of the vibrational mode's ground state wave function in momentum space, $(\Delta p_n)^2 = \hbar m\nu_n/2$, and the corresponding quantity in position space, $(\Delta z_n)^2 = \hbar/2m\nu_n$. Only if $\eta_n$ is nonvanishing, will the absorption or emission of photons be possibly accompanied by a change of the motional state of the atom. Trapping a $^{171}$Yb$^+$ ion, for example, with $\nu_1 = 100 \times 2\pi$kHz gives $\Delta z_1 \approx 17$nm and it is clear from eq. 47 that driving radiation in the optical regime is necessary to couple internal and external dynamics of these trapped ions.

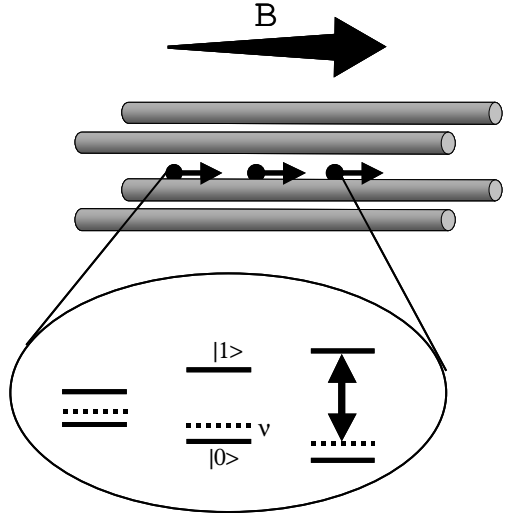

Figure 15: Illustration of a linear ion trap including an axial magnetic field gradient. The static field makes individual ions distinguishable in frequency space by Zeeman-shifting their internal energy levels (solid horizontal lines represent qubit states). In addition, it mediates the coupling between internal and external degrees of freedom when a driving field is applied (dashed horizontal lines stand for vibrational energy levels of the ion string, see text).

### 6.2.2 Spin resonance with trapped ions

As was briefly outlined in the introductory section 1, it would be beneficial for ion trap experiments to take advantage of the highly developed technological resources used in spin resonance (e.g., NMR) experiments. In particular, employing microwave radiation with extremely long coherence time compared to optical radiation allows for precise and, on the time scale of typical experiments, virtually decoherence free manipulation of qubits [2]. In what follows it is outlined how in a linear ion trap with an additional axial magnetic field gradient, $\partial_z B$ i) ions can be individually addressed in frequency space, and ii) the Hamiltonian governing the interaction between microwave radiation and ions is formally identical with 46, with the usual Lamb-Dicke parameter $\eta$ replaced by a new effective LDP $\eta'$ scaling with $\partial_z B/\nu_1^{3/2}$ (Mintert & Wunderlich 2001, Wunderlich 2001).

**Individual addressing of qubits in a modified ion trap** Applying a magnetic field gradient $\vec{B} = bz \cdot \hat{z} + B_0$ along the axial direction of a linear ion trap causes a $z$-dependent Zeeman shift of the internal ionic states $|0\rangle$ and $|1\rangle$. Thus the transition frequency $\omega_{01}^{(j)}$, $j = 1\ldots N$, of each ion is individually shifted and the qubits can be addressed in frequency space. The Breit-Rabi

---

[2]Optical radiation with a long coherence time has been realized experimentally (for instance, (Rafac et al. 2000)). However, building and maintaining such intricate light sources is exceedingly challenging compared to the case of rf or mw radiation

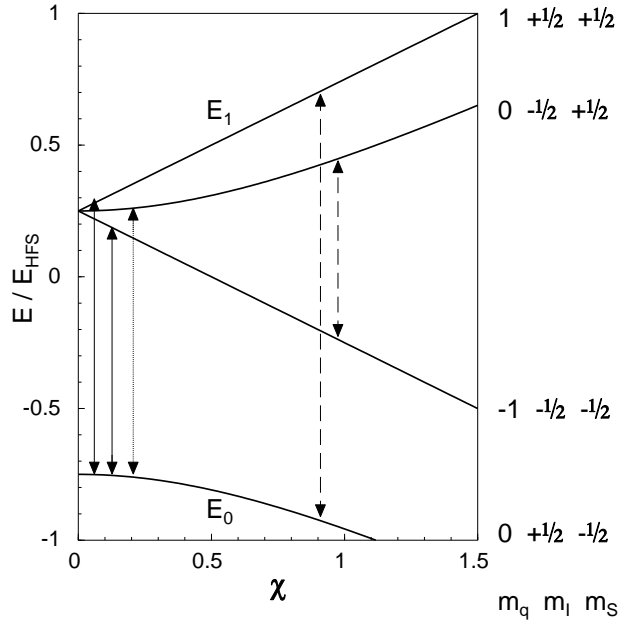

Figure 16: Hyperfine levels of an atom with nuclear spin $I = 1/2$ and electron angular momentum $J = 1/2$ (for Yb$^+$ $J = 1/2 = S$) as a function of scaled magnetic field. Magnetic dipole transitions are indicated for $\pi$−polarized radiation (solid lines, weak field; dashed lines, strong field), and for $\sigma$−polarization (dotted line). The levels marked $E_0$ and $E_1$ are well suited to serve as qubit states.

formula (Corney 1977)

$$E_{m_I m_J} = \frac{E_{\text{HFS}}}{2(2I+1)} - g_I \mu_N B m_q \pm \frac{E_{\text{HFS}}}{2}\left[1 + \frac{4m_q \chi_j}{2I+1} + \chi_j^2\right]^{\frac{1}{2}} \quad (48)$$

gives the energy levels of the hyperfine levels for electron total angular momentum $J = 1/2$ and arbitrary values of the nuclear spin $I$. The hyperfine splitting between levels with total angular momentum $F = I + 1/2$ and $F = I - 1/2$ in zero magnetic field is denoted by $E_{\text{HFS}}$, $m_q = m_I \pm 1/2$, and the plus (minus) sign in front of the last term in 48 is to be used for levels originating from zero-field levels $F = I + 1/2$ ($F = I - 1/2$). The dimensionless quantity $\chi_j$ is defined as

$$\chi_j \equiv \frac{\left(g_J + g_I \frac{m_e}{m_p}\right) \mu_B B(z_j)}{E_{HFS}} \quad (49)$$

where $m_e$ and $m_p$ indicate the electron and proton mass, respectively, $g_J$ and $g_I$ are the electronic and nuclear g-factor, and $\mu_B$ is the Bohr magneton.

Figure 16 shows a plot of the hyperfine levels of the ground state of Yb$^+$ as a function of the scaled magnetic field, $\chi$, and the allowed magnetic dipole transitions are also displayed. In a weak static magnetic field $\vec{B}$ the selection rules $\Delta m_F = \pm 1$ and $\Delta F = 0, \pm 1$ hold for $\pi$-polarized radiation (that is, the electric field vector is parallel to $\vec{B}$; solid lines in Figure 16,) and $\Delta m_F = 0$ and $\Delta F = \pm 1$ are valid for $\sigma$-polarization (dotted line in Figure 16.) In a strong static field the selection rules are $\Delta m_S = \pm 1$ and $\Delta m_I = 0$ for $\pi$-polarized radiation (dashed lines) and $\Delta m_S = 0$ and $\Delta m_I = 0$ for $\sigma$-polarization (no allowed transitions). Therefore, in order to avoid unwanted overlap of resonance frequencies, $E_0$ and $E_1$ are the appropriate choice as qubit states. For the case of the ground state of $^{171}$Yb$^+$ where $E_{\text{HFS}}/\hbar = 12.6 \times 2\pi$GHz, a strong magnetic field (i.e., $\chi \approx 1$) amounts to 0.45T.

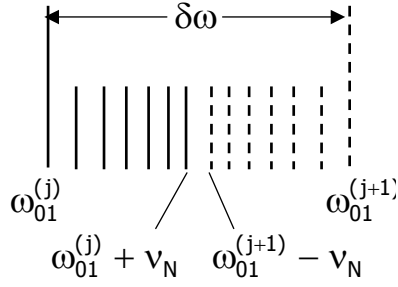

Figure 17: Schematic drawing of the resonances of qubits $j$ and $j+1$ with some accompanying sideband resonances. The angular frequency $\nu_N$ corresponds to the $N$th axial vibrational mode, and the frequency separation between carrier resonances is denoted by $\delta\omega$.

Choosing the levels $E_0$ and $E_1$ indicated in Figure 16 as qubit states and neglecting the contribution of the nuclear spin to the total energy (since the nuclear magneton $\mu_N \ll \mu_B$), the dependence of the qubit resonance frequency on the axial coordinate is given by

$$\frac{\partial \omega_{01}^{(j)}}{\partial z} = \frac{1}{2\hbar} g_J \mu_B \frac{\partial B(z)}{\partial z} \left(1 + \frac{\chi_j}{\sqrt{1+\chi_j^2}}\right) . \qquad (50)$$

The electronic angular momentum is due to the spin of a valence electron and we have $g_J = g_S = 2$.

When separating the qubit resonance frequencies through the application of a magnetic field gradient, overlap between the motional sidebands of the qubit transitions has to be avoided. Therefore, the gradient has to be chosen such that

$$\delta\omega \geq 2\nu_N + \nu_1 \qquad (51)$$

where $\delta\omega = \frac{\partial \omega_{01}}{\partial z} \delta z$ is the frequency shift between two neighboring ions (compare Figure 17), and $\nu_N$ is the angular frequency of the highest axial vibrational mode. Together with expression 44 giving the distance between two ions, $\delta z$, the requirement 51 leads to an estimate of the necessary field gradient in the weak field limit:

$$\frac{\partial B}{\partial z} \geq \frac{\hbar}{2\mu_B} \left(\frac{4\pi\varepsilon_0 m}{e^2}\right)^{\frac{1}{3}} \nu_1^{\frac{5}{3}} \left(4.7 N^{0.56} + 0.5 N^{1.56}\right) . \qquad (52)$$

Thus, for example, $N = 10$ $^{171}$Yb$^+$ ions with $\nu_1 = 100 \times 2\pi$kHz require $\frac{\partial B}{\partial z} \approx 10$ T/m (Mintert & Wunderlich 2001).

The expression 52 for the required field gradient gives an order of magnitude estimate that is necessary to assess whether the necessary gradients are feasible. The exact magnitude of the field gradient has to be determined individually for a given experimental situation in order to also avoid possible interference from second order (in $\eta'$) motional sideband resonances. If, for example, 10 Yb$^+$ ions are used, a constant gradient can be chosen such that it leads to a frequency shift between neighboring ions of $8.8\nu_1$ (while equation 52 yields a gradient equivalent to a frequency shift $\geq 8.6\nu_1$.) Then all second order resonances are separated from the carrier and the respective upper and lower sidebands by at least $0.2$ $\nu_1$. Since the distance between neighboring ions depends on the position of two ions in the linear string, not all ions' resonances will be centered in the desired frequency gap for a constant field gradient. This can be corrected by a slight variation of the gradient along the trap axis (which can be achieved when current carrying coils are used.) Note that the local variation of the field gradient over the extent of the spatial wave function of an individual laser cooled ion would still essentially be zero. Simply increasing the field gradient given in 52 by a factor 2 removes all possible coincidences of first order and second order resonances. Resonances of order three or higher in the effective Lamb-Dicke parameter $\eta'$ possibly still coincide with the useful ones. However their excitation will be suppressed by at least a factor $(\eta')^2$ compared to the first order resonances.

An example may illustrate how the required gradients can be generated: Using a coil of 1 mm diameter (approximately the size of the ion traps employed for the experimental work described in this article) with 3 windings and running a current of 3.3A through them produces a field gradient up to 20 T/m over the required distance. With additional coils the gradient can be modelled to have a desired spatial dependence. With small permanent magnets gradients of a few hundred T/m are easily generated.

**Coupling internal and external dynamics** In the previous paragraphs it was shown that a magnetic field gradient applied to a linear ion trap allows for individual addressing of ions in frequency space. In order to create entangled states between internal and motional degrees of freedom of one particular ion it is obviously necessary to couple internal and external dynamics. If this coupling is possible, then the qubit state of a particular ion may be "written" into the vibrational motion of the ion string, and in subsequent operations transferred to another qubit, or quantum dynamics of one qubit conditioned on the state of another can be performed. At the beginning of this section, the physical reason for coupling of internal and external states was outlined, if laser light is used to drive an internal resonance. If mw or rf radiation is used, the recoil on the ion upon absorption or emission of a photon is not sufficient to excite motional states of the ion (the LDP is vanishingly small.) However, in the presence of a magnetic field gradient motional quanta can nevertheless be created or annihilated in conjunction with changing the internal state of an ion (Mintert & Wunderlich 2001). The physical origin of this effect will be discussed in what follows.

Figure 18 displays two internal states of an ion and a phase space diagram of a harmonic oscillator (an eigenmode of the ion string.) The internal states of the ion are, in the presence of an axial magnetic field gradient, Zeeman shifted as a function of position along the axial direction. In Figure 18 the derivative of the Zeeman shift with respect to the magnetic field has a different sign for the lower energy state $|0\rangle$ and the upper state $|1\rangle$. The position dependent Zeeman shift gives rise to a force acting on the ion in addition to the electrodynamic and Coulomb potentials such that its equilibrium position is slightly different, depending on whether it is in state $|0\rangle$ or $|1\rangle$. Consequently, if an electromagnetic field is applied to drive this qubit resonance, a transition between the two states $|0\rangle$ and $|1\rangle$ will be accompanied by a change of the equilibrium position of the ion,

$$d_z^{(nj)} = -\hbar \frac{\partial_z \omega_{01}^{(j)}}{m\nu_n^2} \quad . \tag{53}$$

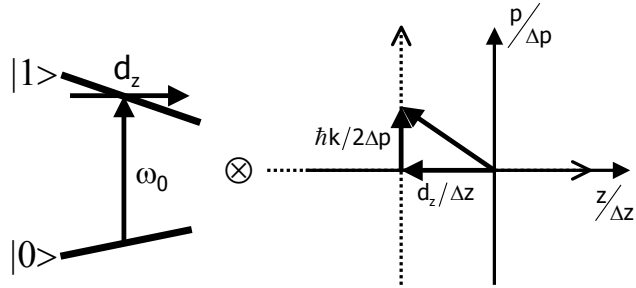

Figure 18: Illustration of the coupled system 'qubit ⊗ harmonic oscillator' in a trap with magnetic field gradient. Internal qubit transitions lead to a displacement $d_z$ of the ion from its initial equilibrium position and consequently to the excitation of vibrational motion. In the formal description the usual Lamb-Dicke parameter is replaced by a new effective one (see text.)

In the phase space diagram of the harmonic oscillator this gives rise to a corresponding shift along the position coordinate together with a shift along the momentum coordinate. (The latter, however, is negligibly small in the microwave regime and is exaggerated in the sketch in Figureo 18.) Thus, the oscillator will be excited and will oscillate about its new equilibrium position. In (Mintert & Wunderlich 2001) it is shown that the formal description of this coupling between internal and external dynamics is identical to the one used for the coupling induced by optical radiation (equation 46.) The usual LDP is replaced by an effective new parameter

$$\eta'_{nj} e^{i\phi_j} \equiv \eta_n S_{nj} + i\epsilon_{nj} \quad . \tag{54}$$

When using mw radiation and appropriately choosing the trap parameters secular axial frequency $\nu_1$ and magnetic field gradient, then $\eta_n S_{nj} \ll \epsilon_{nj}$ and we have $\eta_{nj} \approx \epsilon_{nj}$ with

$$\epsilon_{nj} \equiv S_{nj} \frac{-d_z^{(nj)}}{\Delta z_n} = S_{nj} \frac{\Delta z_n \partial_z \omega_{01}^{(j)}}{\nu_n} \quad . \tag{55}$$

The numerator on the rhs of equation 55 contains the spatial derivative of the resonance frequency of qubit $j$ times the extension $\Delta z_n$ of the ground state wave function of mode $n$, that is, the variation of the internal transition frequency of qubit $j$ when it is moved by a distance $\Delta z$. Thus, the coupling constant $\epsilon_{nj}$ is proportional to the ratio between this frequency variation and the frequency of vibrational mode $n$. The strength of the coupling between an ion's internal dynamics and the motion of the ion string is different for each ion $j$ and depends on the vibrational mode $n$: $S_{nj}$ is a measure for how much ion $j$ participates in the motion of vibrational mode $n$.

All optical schemes devised for conditional quantum dynamics with trapped ions can also be applied in the microwave regime, despite the negligible recoil associated with this type of radiation. This includes, for instance, the proposal presented in (Cirac & Zoller 1995) that requires cooling to the motional ground state, and the proposals reported in (Sorensen & Molmer 2000) and (Jonathan & Plenio 2001) (the latter two work also with ions in thermal motion.)

**Trapped ions as a $N$-qubit molecule**   The Hamiltonian describing a string of trapped two-level ions in a trap with axial magnetic field gradient (without additional radiation used to drive internal transitions) has been shown to read (Wunderlich 2001)

$$H = \frac{\hbar}{2}\sum_{j=1}^{N}\omega_j(z_{0,j})\sigma_{z,j} + \sum_{n=1}^{N}\hbar\nu_n(a_n^\dagger a_n) - \frac{\hbar}{2}\sum_{i<j}^{N}J_{ij}\sigma_{z,i}\sigma_{z,j} \; . \qquad (56)$$

The first sum on the rhs of 56 represents the internal energy of the collection of $N$ ions. The qubit angular resonance of ion $j$ at its equilibrium position $z_{0,j}$ is $\omega_j$. The second term sums the energy of $N$ axial vibrational modes. These first two terms represent the usual Hamiltonian for a string of two-level ions confined in a harmonic potential (James 1998). The new spin-spin coupling term (last sum in 56) arises due to the presence of the magnetic field gradient. Here,

$$J_{ij} \equiv \sum_{n=1}^{N}\nu_n\epsilon_{ni}\epsilon_{nj} \; . \qquad (57)$$

The pairwise coupling 57 between qubits $i$ and $j$ is mediated by the vibrational motion. Therefore, it contains terms quadratic in $\epsilon$, and the coupling of qubit $i$ and $j$ to the vibrational motion has to be summed over all modes.

As an example, Table 1 shows the spin-spin coupling constants between 10 $^{171}$Yb$^+$ ions confined in a linear trap ($\nu_1 = 100 \times 2\pi$kHz) with a magnetic field gradient of 25T/m. The application of NMR-type quantum logic operations to such an artificial molecule is facilitated by the fact that individual qubit resonances are widely separated in frequency (in this example, the frequency "gap between neighboring ions is about 1 MHz) as compared to typical NMR experiments (Vandersypen et al. 2001). In addition, the coupling constants $J_{ij}$ have similar and nonzero values for all pairs of spins.

In a "real" molecule different nuclear spins share binding electrons that generate a magnetic field at the location of the nuclei, and the energy of a nuclear spin exposed to the electrons' magnetic field depends on the charge distribution of the binding electrons. If a particular nuclear spin is flipped, the interaction with the surrounding electrons will slightly change the electrons' charge distribution which in turn may affect the energy of other nuclear spins. This indirect spin-spin coupling is realized here in a different way: the role of the electrons' magnetic field is replaced by the vibrational motion of the ions.

| $i$ | $J_{i1}$ | $J_{i2}$ | $J_{i3}$ | $J_{i4}$ | $J_{i5}$ | $J_{i6}$ | $J_{i7}$ | $J_{i8}$ | $J_{i9}$ |
|---|---|---|---|---|---|---|---|---|---|
| 1 | | | | | | | | | |
| 2 | 54.61 | | | | | | | | |
| 3 | 41.36 | 48.12 | | | | | | | |
| 4 | 34.15 | 38.89 | 44.74 | | | | | | |
| 5 | 29.40 | 33.17 | 37.44 | 43.04 | | | | | |
| 6 | 25.92 | 29.09 | 32.55 | 36.77 | 42.52 | | | | |
| 7 | 23.19 | 25.93 | 28.88 | 32.35 | 36.77 | 43.04 | | | |
| 8 | 20.92 | 23.33 | 25.90 | 28.88 | 32.55 | 37.44 | 44.74 | | |
| 9 | 18.93 | 21.07 | 23.33 | 25.93 | 29.09 | 33.17 | 38.89 | 48.12 | |
| 10 | 17.04 | 18.93 | 20.92 | 23.19 | 25.92 | 29.40 | 34.15 | 41.36 | 54.61 |

Table 1: Spin-spin coupling constants $J_{ij}/2\pi$ in units of Hz for 10 $^{171}$Yb$^+$ ions in a linear trap characterized by the angular frequency of the COM vibrational mode $\nu_1 = 100 \times 2\pi$kHz using a static field gradient of 25T/m.

Usual ion trap schemes take advantage of motional sidebands that accompany qubit transitions. Instead, the spin-spin coupling that arises in a suitably modified trap may be directly used to implement conditional dynamics using NMR methods. The collection of trapped ions can thus be viewed as a $N$-qubit molecule with adjustable coupling constants (Wunderlich 2001). Making use of this spin-spin coupling does not involve real excitation of vibrational motion. In this sense it is similar to a scheme for conditional quantum dynamics that uses optical 2-photon transitions detuned from vibrational resonances (Sorensen & Molmer 2000), and, thus should be tolerant against thermal motion of the ions.

## 6.3 Coherent optical excitation with Ba$^+$ and Yb$^+$ ions

This section is devoted to another possible, more "traditional" avenue towards quantum computation with trapped ions: employing an optical transition as a qubit. Since the relaxation rates of the states acting as a qubit eventually limit the time available for coherent manipulation, the use of two states connected via an electric dipole allowed transition is not a good choice. Therefore, the electronic ground state of an ion (usually one specific Zeeman sublevel) and a metastable excited state have been chosen as qubit states in various experiments (Appasamy et al. 1998, Roos et al. 1999, Barton et al. 2000). Here, we report on experiments with Ba$^+$ and $^{172}$Yb$^+$ ions where the electric quadrupole (E2) resonance $^2$S$_{1/2}$-$^2$D$_{5/2}$ serves as the qubit.

### 6.3.1 Rabi oscillations on optical E2 resonance in Ba$^+$

The lifetime of the metastable state $^2$D$_{5/2}$ ($\approx 34$s) is long on the timescale of typical coherent operations on this transition. Thus, the useful coherence time for quantum logic operations with this qubit transition is essentially limited i) by the inverse emission bandwidth of laser light close to 1.76$\mu$m driving the E2 resonance, and ii) by the stability of static magnetic fields that lift the degeneracy of Zeeman states. Exciting sideband resonances of this E2 transition allows for coupling of internal and external degrees of freedom (compare section 6.2.)

Ba$^+$-ions are confined in a 1-mm-diameter Paul trap and irradiated by laser light at 493 nm (green light) for excitation of resonance fluorescence on the $^2$S$_{1/2}$-$^2$P$_{1/2}$ transition (compare Figure 19.) This laser is usually detuned a few ten MHz below resonance for cooling the ions. Tunable light close to 493nm is obtained by first generating light near 986 nm by a diode laser (stabilized to a reference resonator), and then frequency doubling the infrared light in a ring resonator containing a KNbO$_3$ crystal as a nonlinear element. A dye laser at 650 nm (red light) prevents optical pumping into the $^2$D$_{3/2}$ level. The fluorescence signal is recorded by photon counting. A static magnetic field defines the quantization axis and lifts the degeneracy of the magnetic sublevels. The direction of propagation and the polarization of both light beams (green and red) are set perpendicular to the magnetic field. The power levels of the light fields are stabilized by electro-optic modulators.

The attainable Rabi frequency on the optical E2 resonance S$_{1/2}$ - D$_{5/2}$ near 1762nm (having a spectral width of 5mHz) is limited by the available intensity of the light exciting this transition, and by the emission bandwidth, $\Delta\omega$ of the laser light. In addition, $\Delta\omega$ determines the coherence time of the qubit. A color-center laser that delivered up to 150mW of light near 1762nm with an effective emission bandwidth $\Delta\omega \approx 30 \times 2\pi$kHz ($1/e^2$ full width of a Gaussian profile) was previously in use to excite this resonance (Appasamy et al. 1998). The considerable effort that has to be devoted to the preparation of suitable NaCl crystals, the necessity to always maintain the laser medium at liquid nitrogen temperature, and the need for an Argon ion laser that ensures the correct polarization of the color-centers, in addition to the pump laser at 1064nm are but a few of the obstacles that make such a laser a time consuming and not very economical instrument. This laser was replaced by a continuous wave optical parametric oscillator (Linos AG) emitting light in the required wavelength range. In order to attain the desired long term stability of the emission frequency, a highly stable reference resonator suspended in ultra-high vacuum was used. Insensitivity against vibrations and variations in temperature and air pressure is thus ensured (Leick 2000).

Figure 20 shows Rabi oscillations on the carrier transition of the E2-resonance in Ba$^+$ . Each data point is obtained by executing the following sequence 600 times: i) the infrared light driving the E2 transition is switched on for a time $\tau$ indicated on the abscissa in Figure 20 while the green light exciting the dipole resonance S$_{1/2}$-P$_{1/2}$ is turned off. ii) Laser light near 493nm is turned on for 1ms, and scattered light is collected during this time for state selective detection: Either scattered photons will be detected during the last step, indicating that the state of the ion is S$_{1/2}$ at the end of step ii (the registered number of photons is Poisson distributed around a mean value of typically 10 counts.) Or, if no photon counts are registered, the ion was in state D$_{5/2}$. Thus, a trajectory of "on" (resonance fluorescence is observed) and "off" (absence of resonance fluorescence) events is recorded. A pair of "on"-"off" events indicates a transition from state S$_{1/2}$ to state D$_{5/2}$. The probability for absorption of an ir photon is calculated by dividing the number of these excitation events by the number of "on" events (total number of tries of excitation) in a trajectory. The probability for emission is obtained analogously. In Figure 20 absorption and emission probability from one trajectory for a given time $\tau$ have been averaged to yield the transition probability between states $S_{1/2}$ and $D_{5/2}$. A fit of the data displayed in Figure 20 yields a Rabi frequency of $71.4 \times 2\pi$kHz and a transversal relaxation time of $100\mu$s. Dephasing is determined by the emission bandwidth of the ir laser which will be further narrowed by improved frequency locking of the ir laser in future experiments.

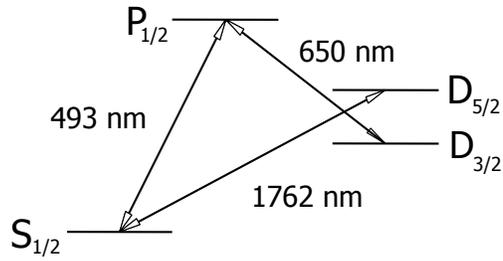

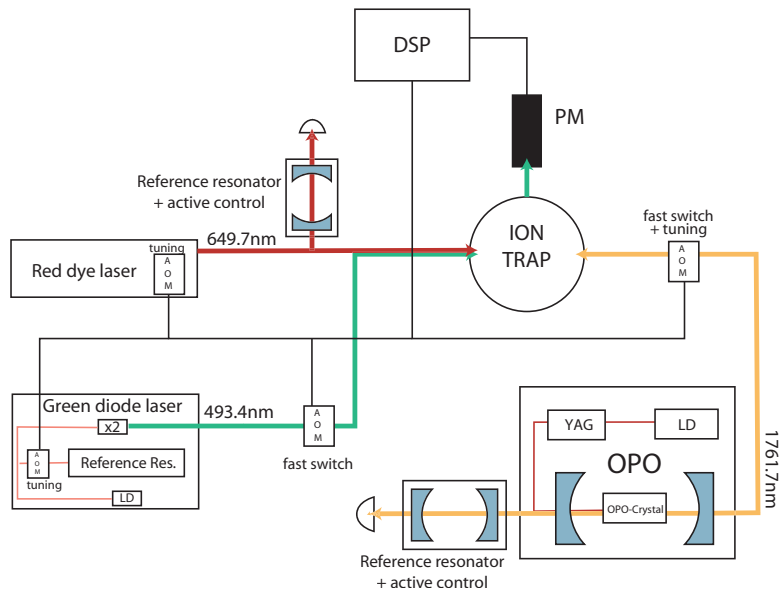

Figure 19: a) Relevant energy levels and transitions in $^{138}$Ba$^+$. b) Schematic drawing of major experimental elements. OPO: Optical parametric oscillator; YAG: Nd:YAG laser; LD: laser diode; DSP: Digital signal processing system allows for real time control of experimental parameters; AOM: Acousto-optic modulators used as optical switches and for tuning of laser light; PM: Photo multiplier tube, serves for detection of resonance fluorescence. All lasers are frequency and intensity stabilized (not shown.)

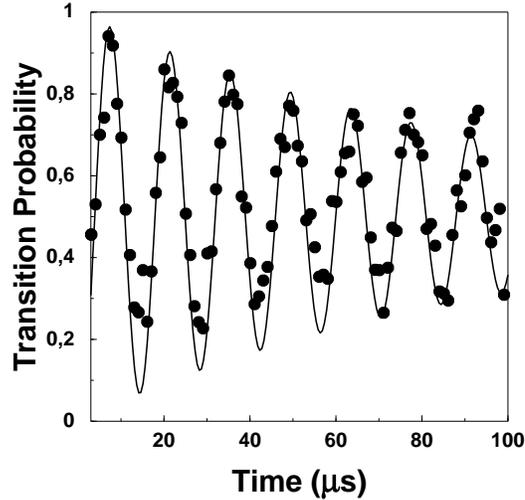

Figure 20: Rabi oscillations on the optical E2 transition $S_{1/2}$-$D_{5/2}$ in $Ba^+$ . A fit of the data (solid line) yields a Rabi frequency of $71.4 \times 2\pi$kHz and a transversal relaxation time of $100\mu$s (determined by the coherence time of the ir light used to drive the E2 resonance.)

### 6.3.2 Lifetime measurement of the $D_{5/2}$ state in $Ba^+$

When using the metastable $D_{5/2}$ state in $Ba^+$ as one quantum state of a qubit, or for the potential application of electrodynamically trapped $Ba^+$ as a frequency standard, it is useful to know the lifetime of this state. $Ba^+$ is also a promising candidate to measure parity nonconserving interactions in atoms complementing high energy experiments in search of new physics beyond the standard model (Fortson 1993, Geetha et al. 1998). Comparison between results obtained from atomic structure theory and experimentally determined values are thus important. Previous attempts of determining the lifetime of the $D_{5/2}$ state have yielded different values in experiments with single and many ions, respectively (Plumelle et al. 1980, Nagourney et al. 1986, Madej & Sankey 1990). In recent experiments on $Ca^+$ it was found that the lifetime of the $D_{5/2}$ state in this ion species depended on the power of an additional laser used to repump the ion from the metastable $D_{3/2}$ state (Block et al. 1999).

We have determined the lifetime of the $D_{5/2}$ state in $Ba^+$ using the quantum jump method. The resulting experimental lifetime is limited by collisions with background gas to 21 s which agrees well with the results reported in (Madej

& Sankey 1990). We did not find a dependence on the power or detuning of the laser used to scatter resonant light on the S$_{1/2}$-P$_{1/2}$ transition or of the "repumper" from the D$_{3/2}$ state.

### 6.3.3 Cooling of a pair of Ba$^+$ ions

Cooling of the collective motion of several particles, not necessarily to the motional ground state (Sorensen & Molmer 2000, Jonathan & Plenio 2001) is prerequisite for implementing conditional quantum dynamics with trapped ions. We have studied the collective vibrational motion of two trapped $^{138}$Ba$^+$ ions cooled by laser light close to the resonances corresponding to the S$_{1/2}$-P$_{1/2}$ (493 nm, green light) and P$_{1/2}$-D$_{3/2}$ (650 nm, red light) transition, respectively.

When two ions are confined in a nearly spherically symmetric Paul trap, and if they are sufficiently laser cooled, then we always observe them crystallizing at the same locations. The crystallization at preferred locations is explained by the slight asymmetry of the effective trapping potential, that is, $\nu_x \neq \nu_y \neq \nu_z$, where $\nu_{x,y,z}$ are the angular frequencies of the center-of-mass-mode of the secular motion in different spatial directions. The Coulomb potential makes the ions repel each other, and the ion crystal tends to align along the axis of weakest confinement by the electrodynamic potential.

The potential along the $z-$direction is steeper than in the $xy$- plane. Consequently, if cooled well enough, the ions stay in this plane. Since $\nu_y > \nu_x$ they are not free to rotate in the xy-plane, and instead would have to surmount an azimuthal potential barrier at $\phi = \pi/2$ ($\phi = \arctan(y/x)$) in order to exchange places. However, if the vibrational energy of the relative motion of the two ions in the $\tilde{y}$-mode (Reiß et al. 2002) exceeds the azimuthal barrier height, then the ions are free to rotate in the $xy-$plane. Depending on the parameter settings (detuning and intensity) of the cooling lasers, these different motional states corresponding to different temperatures are indeed observed experimentally.

If, for instance, the intensity and frequency of the green laser is held fixed and the red laser's frequency is scanned, then a characteristic spectrum displaying dark resonances is obtained (Figure 21). Whenever the detuning of the red laser, with respect to a resonance between a Zeeman level of the D$_{3/2}$ state and one of the P$_{1/2}$ state equals the detuning of the green laser with respect to a resonance between Zeeman levels of the P$_{1/2}$ and S$_{1/2}$ states, a coherent superposition of the Zeeman levels of S$_{1/2}$ and D$_{3/2}$ is created that does not couple to the light field anymore. The appearance of four dark resonances is due to the selection rules for dipole allowed transitions between Zeeman sublevels of the S$_{1/2}$, P$_{1/2}$, and D$_{3/2}$ electronic states when both light fields are linearly polarized perpendicular to the magnetic field that defines the quantization axis. Fitting such an excitation spectrum using the optical Bloch equations allows for the determination of intensity and detuning of the laser light, as well as of the strength of the applied magnetic field. Upon scanning the red laser it is observed that ions take on different states of motion: either they crystallize at fixed locations or they form a ring-shaped structure when their thermal energy is sufficient to surmount the azimuthal potential barrier.

Using the laser parameters determined from a fit of the excitation spectrum, the expected temperature of the ions can be derived from detailed numerical calculations of laser cooling taking into account the Zeeman structure of the energy levels (Reiß et al. 2002). It turns out that the transition from an ion crystal to the ring structure occurs at that detunings of the red laser where theory predicts laser cooling to turn into heating. The ions gain enough energy from scattering photons to surmount the azimuthal potential barrier and appear as a ring on the spatially resolving photo detector. The transition from cooling to heating occurs when the red laser is scanned across a dark resonance with increasing frequency: as soon as it is blue detuned with respect to the closest dark resonance, the cooling rate is reduced to zero and with further increasing laser frequency becomes negative (that is, heating occurs.) Increasing the laser's frequency even more means that the red laser is further blue detuned with respect to the dark resonance that was just passed. At the same time, however, the next resonance is approached relative to which the laser is red detuned and the cooling rate increases again. It should be noted that Raman scattering responsible for these processes occurs when both laser are *red* detuned relative to the main resonance.

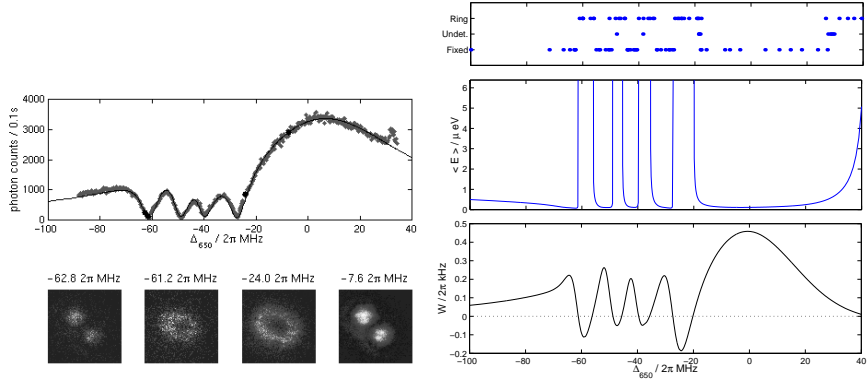

Figure 21: Left top: Fluorescence of two trapped ions as a function of laser detuning. Left bottom: Two trapped Ba$^+$ ions show different motional states depending on laser parameters. Spatial distribution of the two ions at the detunings indicated above. Right top: Observed motional states for different detunings of the 650 nm light. The dots correspond to individual observations. Right middle: Mean motional energy in the $\tilde{y}$-mode calculated from theory. Right bottom: Cooling rate for the $\tilde{y}$-mode calculated from theory. Figure taken from (Reiß et al. 2002).

Very good agreement is found between the theoretical prediction of the transition of the ions' motional state and experimental observations. In addition, parameter regimes of the laser light irradiating the ions are identified that imply most efficient laser cooling and are least susceptible to drifts, fluctuations, and uncertainties in laser parameters. When applied to cooling of a string of ions in a linear trap, the multidimensional parameter space allows to identify regions where cooling is most efficient for all vibrational modes. In particular, the magnetic field can be increased for a larger separation of the dark resonances.

Cooling of different vibrational modes is also achieved with electromagnetically induced transparency (EIT) cooling (Morigi et al. 2000, Roos et al. 2000). In that scheme, too, atomic resonances are shaped by two laser fields such that most efficient cooling for as many vibrational modes as possible is achieved.

### 6.3.4 Coherent excitation of an E2 resonance in $^{172}$Yb$^+$

The electric quadrupole resonance $S_{1/2} \leftrightarrow D_{5/2}$ in $^{172}$Yb$^+$ with a natural linewidth of $6 \times 2\pi$Hz (Fawcett & Wilson 1991) may be used as a qubit, too. The relevant energy levels involved in the investigation of coherent excitation of this transition are shown in Figure 1. The E2-transition is driven by light of a frequency doubled diode laser at 411nm. The population of the $S_{1/2}$ ground state is probed by exciting resonance fluorescence on the strong monitor transition $S_{1/2} \leftrightarrow P_{1/2}$. In addition to spontaneous decay into the $S_{1/2}$ ground state, the state $D_{5/2}$ might decay into the extremely long lived level $F_{7/2}$ (lifetime $> 10$ years (Roberts et al. 1997)) with probability 0.81. The population trapped in the $F_{7/2}$ state is brought back into the $S_{1/2}$ ground state via the jk-coupled (Cowan 2001) level $D[5/2]_{5/2}$ by illuminating the ion continuously with laser light at 638nm. The depopulation time depends on the laser intensity and is found to be $\tau_{638} = 9$ms in the limit of high intensity.

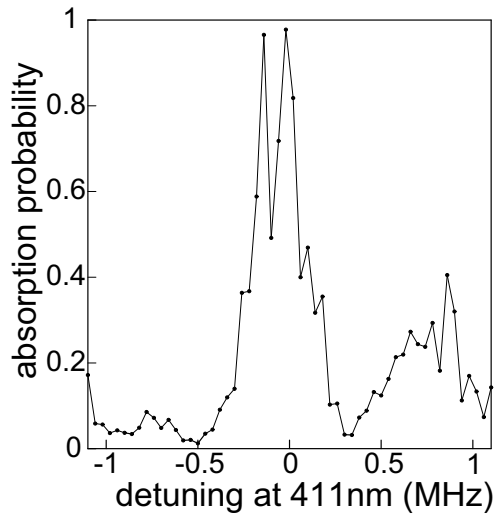

Figure 22: Absorption spectrum between the Zeeman states $|S_{1/2}, m_j = 1/2\rangle \leftrightarrow |D_{5/2}, m_j = 1/2\rangle$ (E2-transition.) The absorption probability on resonance exceeds 0.9 which proves coherent excitation. The sidebands at $\pm 750$kHz next to the carrier are caused by the secular motion of the ion.

An absorption spectrum of the E2 transition is obtained by scanning the frequency of the light at 411nm in steps of 40kHz across the resonance of a selected Zeeman component. At every frequency step a series of 500 pairs of driving pulses ($\tau_{411} = 5$ms) and probing pulses ($\tau_{369} = 10$ms) is recorded resulting in a trajectory of "on" and "off" observations. The absorption probability on the $|S_{1/2}, m_j = 1/2\rangle \leftrightarrow |D_{5/2}, m_j = 1/2\rangle$ transition, determined in the same way as described in the previous section, is plotted in Figure 22 versus the detuning of the frequency of the light field at 411nm.

The measured absorption probability on the carrier transition exceeds 0.9 verifying coherent excitation of the E2 resonance. The structure seen in the carrier is due to Rabi oscillations (Balzer et al. 2000, Wunderlich et al. 2001).

From the width of the carrier resonance the Rabi frequency is estimated to be ≈ 110 kHz. A comparison of the experimental spectrum with numerical simulations using optical Bloch equations shows that the emission bandwidth of the laser field at 411nm is less than 5Hz in 5ms.

Next to the carrier two sidebands are visible at ±750kHz arising from the ion's axial secular motion in the pseudo harmonic potential of the electrodynamic trap. The asymmetry in the absorption probability between upper and lower sideband is due to sideband cooling on the E2 transition (to be detailed elsewhere.)

Employing optical E2 transitions, too, important steps towards quantum information processing have been experimentally realized. Because of the simple level structure of $^{138}$Ba$^+$, and of the long lifetime of its metastable D$_{5/2}$ state, this ion is well suited for experiments where coherent optical excitation is desired (for instance, in QIP).